\documentclass[a4paper]{article}
\usepackage{graphics,setspace,amsmath}
\usepackage{natbib}

\newenvironment{quoting}{\begin{quote}\singlespacing}{\end{quote}}

\newcommand{\ket}[2]{|#1\rangle _{#2}}
\newcommand{\bra}[2]{\langle _{#1}#2|}
\newcommand{\up}{\!\uparrow}
\newcommand{\down}{\!\downarrow}

\title{The Grammar of Teleportation}
\author{Christopher G. Timpson\thanks{\texttt{c.g.timpson@leeds.ac.uk}} \\ \textit{\small Division of History and Philosophy of Science,} \\ \textit{\small School of Philosophy, University of Leeds,} \\ \textit{\small Leeds, LS2 9JT, UK} }

\begin{document}
\maketitle

\begin{abstract}\noindent Whilst a straightforward consequence of the formalism of non-relativistic quantum mechanics, the phenomenon of quantum teleportation has given rise to considerable puzzlement. In this paper, the teleportation protocol is reviewed and these puzzles dispelled. It is suggested that they arise from two primary sources: 1) the familiar error of hypostatizing an abstract noun (in this case, `information') and 2) failure to differentiate interpretation dependent from interpretation \textit{in}dependent features of quantum mechanics. A subsidiary source of error, the \textit{simulation fallacy}, is also identified. The resolution presented of the puzzles of teleportation illustrates the benefits of paying due attention to the logical status of `information' as an abstract noun.
\end{abstract}

\section{Introduction}


\begin{quoting}
`The questions ``What is length?", ``What is meaning?", ``What is the number one?" etc. produce in us a mental cramp.
We feel that we can't point to anything in reply to them and yet ought to point to something.
(We are up against one of the great sources of philosophical bewilderment: a substantive makes us look for a thing that corresponds to it.)' \citet{wittgenstein:blue}

\end{quoting}

Quantum teleportation (\citet{teleportation}) is one of the singular fruits of the burgeoning field of quantum information theory. In this theory, one seeks to describe and make use of the distinctive possibilities for information processing and communication that quantum systems allow: quantum features like entanglement and non-commutativity are put to work. 

Teleportation is perhaps the most striking example of the use of entanglement in assisting communication and it illustrates vividly several of the general features associated with quantum information protocols, most notably the fact that entanglement (a characteristically quantum property) serves as an important resource, and that unknown quantum states cannot be cloned (\citet{dieks, wootters:zurek}). 

Although a straightforward consequence of the formalism of non-relativistic quantum mechanics, teleportation has given rise to some confusion and to a good deal of controversy.  
In this paper I review the main lines of controversy (Sections \ref{protocol} and \ref{puzzles}) and seek to dispell the confusion that has surrounded the interpretation of the protocol.

I will suggest (Section~\ref{dissolving}) that puzzlement has generally arisen as a consequence of a familiar philosophical error---in fact the one that Wittgenstein famously warns us of in the \textit{Blue Book}---that is, the error of assuming that every grammatical substantive is a referring term. Here the culprit is the word `information'. `Information' is an abstract (mass) noun and hence does not refer to a spatio-temporal particular, to an entity or a substance\footnote{For elaboration of this claim, see \citet[Chpt.1]{thesis}. The point is also made in \citet{nifpaper}.}. It follows that one should not be seeking in an information theoretic protocol---quantum or otherwise---for some particular, denoted by `the information', whose path one is to follow, but rather concentrating on the physical processes by which the information is transmitted, that is, by which the end result of the protocol is brought about. Once this is recognised, I suggest, much of our confusion is dispelled. (A subsidiary source of difficulty---what I term the \textit{simulation fallacy}---is also remarked upon.)

With this clarification in place, the other major source of controversy is thrown into relief: just what \textit{are} the physical processes by which teleportation is effected? This is, in fact, a relatively straightforward question; but it is a question that will find a different answer depending on what interpretation of quantum mechanics one wishes to adopt (Section~\ref{interpretations}) a point which has not been sufficiently recognised to date.

The central theme of this paper is that the conceptual puzzles surrounding teleportation arise from thinking about information in the wrong way. The converse point holds too: the clarification of these puzzles clearly illustrates the value of recognising the logico-grammatical status of `information' as an abstract noun.      


Let us begin by briefly reviewing the teleportation protocol\footnote{Helpful discussions of further conceptual aspects of teleporation, in particular concerning the relation of teleportation to nonlocality, may be found in \citet{hardy:disentangling}, \citet{jon1} and \citet{cliftonpope}. \citet{mermin:teleportation} also provides an interesting perspective.}.

\section{The quantum teleportation protocol}\label{protocol}

In the teleportation protocol we consider two parties, Alice and Bob, who are widely separated, but each of whom possess one member of a pair of particles in a maximally entangled state. Alice is presented with a system in some unknown quantum state, and her aim is to transmit this state to Bob. In the standard example, Alice and Bob share one of the four Bell states (Table~\ref{bellstates}) and she is presented with a spin-1/2 system in the unknown state $\ket{\chi}{}=\alpha\ket{\up}{}+\beta\ket{\down}{}$.

\begin{table}
\begin{center}
\[ \begin{array}{c}
\ket{\phi^{+}}{}=1/\sqrt{2}(\ket{\up}{}\ket{\up}{}+\ket{\down}{}\ket{\down}{}),\\
\ket{\phi^{-}}{}=1/\sqrt{2}(\ket{\up}{}\ket{\up}{}-\ket{\down}{}\ket{\down}{}),\\
\ket{\psi^{+}}{}=1/\sqrt{2}(\ket{\up}{}\ket{\down}{}+\ket{\down}{}\ket{\up}{}),\\
\ket{\psi^{-}}{}=1/\sqrt{2}(\ket{\up}{}\ket{\down}{}-\ket{\down}{}\ket{\up}{}).
\end{array} \]
\end{center}
\caption{The four Bell states, a maximally entangled basis for $2\otimes2$ dimensional systems.}\label{bellstates}
\end{table}

\begin{figure}
\includegraphics{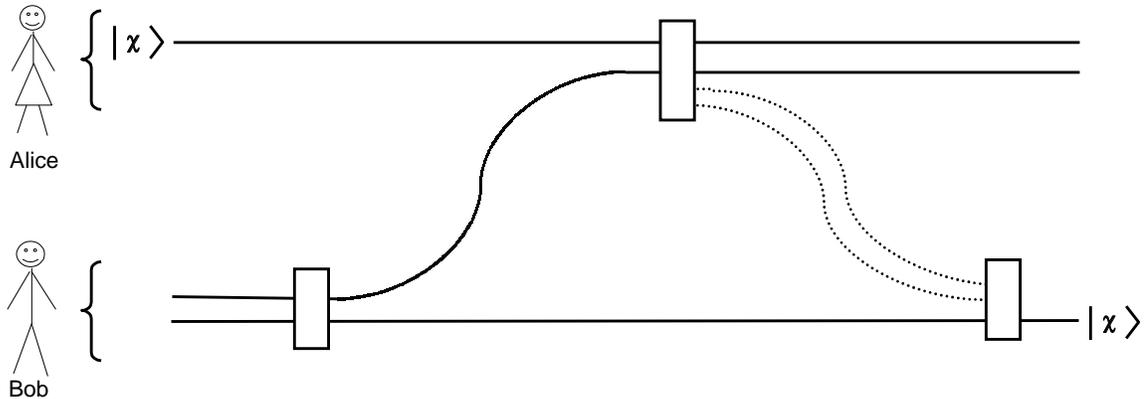}
\caption{\small Teleportation. A pair of systems is first prepared in an entangled state and shared between Alice and Bob, who are widely spatially separated. Alice also possesses a system in an unknown state $\ket{\chi}{}$. Once Alice performs her Bell-basis measurement, two classical bits recording the outcome are sent to Bob, who may then perform the required conditional operation to obtain a system in the unknown state $\ket{\chi}{}$. (Continuous black lines represent qubits, dotted lines represent classical bits. Time runs along the horizontal axis.)\label{telep1}}
\end{figure}

By performing a suitable joint measurement on her half of the entangled pair and the system whose state she is trying to transmit (in this example, a measurement in the Bell basis), Alice can flip the state of Bob's half of the entangled pair into a state that differs from $\ket{\chi}{}$ by one of four unitary transformations, depending on what the outcome of her measurement was. If a record of the outcome of Alice's measurement is then sent to Bob, he may perform the required operation to obtain a system in the state Alice was trying to send (Fig.~{\ref{telep1}}). 

The end result of the protocol is that Bob obtains a system in the state $\ket{\chi}{}$, with nothing that bears any relation to the identity of this state having traversed the space between him and Alice. Only two classical bits recording the outcome of Alice's measurement were sent between them; and the values of these bits are completely random, with no dependence on the parameters $\alpha$ and $\beta$. Meanwhile, no trace of the identity of the unknown state remains in Alice's region, as is required in accordance with the no-cloning theorem (the state of her original system will usually now be maximally mixed). The state has `disappeared' from Alice's region and `reappeared' in Bob's, hence the use of the term \textit{teleportation} for this phenomenon.

To fix the process in our minds, let's review how the standard example goes. We begin with system 1 in the unknown state $\ket{\chi}{}$ and with Alice and Bob sharing a pair of systems (2 and 3) in, say, the singlet state $\ket{\psi^{-}}{}$. The total state of the three systems at the beginning of the protocol is therefore simply
\begin{equation}\label{start}
\ket{\chi}{1}\ket{\psi^{-}}{23} = \frac{1}{\sqrt 2} \bigl(\alpha \ket{\up}{1} + \beta\ket{\down}{1}\bigr)\bigl(\ket{\up}{2}\ket{\down}{3}-\ket{\down}{2}\ket{\up}{3}\bigr).
\end{equation} 
Notice that at this stage, the state of system 1 factorises from that of systems 2 and 3; and so in particular, the state of Bob's system is independent of $\alpha$ and $\beta$. We may re-write this initial state in a suggestive manner, though:

\begin{singlespacing}
\begin{align}
\ket{\chi}{1}\ket{\psi^{-}}{23} & = \frac{1}{\sqrt{2}}\biggl(\alpha\ket{\up}{1}\ket{\up}{2}\ket{\down}{3} + \beta\ket{\down}{1}\ket{\up}{2}\ket{\down}{3} - \alpha\ket{\up}{1}\ket{\down}{2}\ket{\up}{3} - \beta\ket{\down}{1}\ket{\down}{2}\ket{\up}{3}\biggr) \\
   \begin{split} & =  
 \frac{1}{2}\biggl(\ket{\phi^{+}}{12} \bigl(\alpha \ket{\down}{3} - \beta\ket{\up}{3}\bigr)+\ket{\phi^{-}}{12} \bigl(\alpha \ket{\down}{3} + \beta\ket{\up}{3}\bigr) \\
 & \phantom{\ket{\phi^{+}}{} }+ \ket{\psi^{+}}{12} \bigl(-\alpha \ket{\up}{3} + \beta\ket{\down}{3}\bigr) + \ket{\psi^{-}}{12} \bigl(-\alpha \ket{\up}{3} - \beta\ket{\down}{3}\bigr)\biggr). 
\end{split} \label{rewrite1}
\end{align}\end{singlespacing}
The basis used is the set \[\{\ket{\phi^{\pm}}{12}\ket{\up}{3},\,\ket{\phi^{\pm}}{12}\ket{\down}{3},\, \ket{\psi^{\pm}}{12}\ket{\up}{3}, \, \ket{\psi^{\pm}}{12}\ket{\down}{3}\},\] that is, 
we have chosen (as we may) to express the total state of systems 1,2 and 3 using an entangled basis for systems 1 and 2, even though these systems are quite independent. But so far, of course, all we have done is re-written the state in a particular way; nothing has changed physically and it is still the case that it is really systems 2 and 3 that are entangled and wholly independent of system 1, in its unknown state.

Looking closely at (\ref{rewrite1}) we notice that the relative states of system 3 with respect to particular Bell basis states for 1 and 2 have a very simple relation to the initial unknown state $\ket{\chi}{}$; they differ from $\ket{\chi}{}$ by one of four local unitary operations:
\begin{multline}\label{rewrite2}
\ket{\chi}{1}\ket{\psi^{-}}{23} = \frac{1}{2}\biggl( \ket{\phi^{+}}{12} \bigl(-i\sigma_{y}^{3}\ket{\chi}{3}\bigr) + \ket{\phi^{-}}{12} \bigl(\sigma_{x}^{3}\ket{\chi}{3}\bigr) \\
 + \ket{\psi^{+}}{12} \bigl(-\sigma_{z}^{3}\ket{\chi}{3}\bigr) + \ket{\psi^{-}}{12} \bigl(-\mathbf{1}^{3}\ket{\chi}{3}\bigr)\biggr),  
\end{multline}
where the $\sigma_{i}^{3}$ are the Pauli operators acting on system 3 and $\mathbf{1}$ is the identity. To re-iterate, though, only system 1 actually depends on $\alpha$ and $\beta$; the state of system 3 at this stage of the protocol (its reduced state, as it is a member of an entangled pair) is simply the maximally mixed $1/2\,\mathbf{1}$.  

Alice is now going to perform a measurement. If she were simply to measure system 1 then nothing of interest would happen---she would obtain some result and affect the state of system 1, but systems 2 and 3 would remain in the same old state $\ket{\psi^{-}}{}$. However, as she has access to both systems 1 and 2, she may instead perform a \textit{joint} measurement, and now things get interesting. In particular, if she measures 1 and 2 in the Bell basis, then after the measurement we will be left with only one of the terms on the right-hand side of eqn.~(\ref{rewrite2}), at random; and this means that Bob's system will have jumped instantaneously into one of the states $-i\sigma_{y}^{3}\ket{\chi}{3},\, \sigma_{x}^{3}\ket{\chi}{3},\, -\sigma_{z}^{3}\ket{\chi}{3}$ or  $-\ket{\chi}{3}$, with equal probability.

But how do things look to Bob? As he neither knows whether Alice has performed her measurement, nor, if she has, what the outcome turned out to be, he will still ascribe the same, original, density operator to his system---the maximally mixed state\footnote{Notice that an equal mixture of the four possible post-measurement states of his system results in the density operator $1/2\, \mathbf{1}$.}. No measurement on his system could yet reveal any dependence on $\alpha$ and $\beta$. To complete the protocol therefore, Alice needs to send Bob a message instructing him which of four unitary operators to apply $(i\sigma_{y},\,\sigma_{x},\,-\sigma_{z},\,\mathbf{-1})$ in order to make his system acquire the state $\ket{\chi}{}$ with certainty; for this she will need to send two bits\footnote{Two bits are clearly sufficient, for the argument that they are strictly necessary, see \citet{teleportation} Fig.2.}. With these bits in hand, Bob applies the needed transformation and obtains a system in the state $\ket{\chi}{}$.

Now of course, this quantum mechanical process differs from science fiction versions of teleportation in at least two ways. First, it is not \textit{matter} that is transported, but simply the quantum state $\ket{\chi}{}$; and second, the protocol is not instantaneous, but must attend for its completion on the arrival of the classical bits sent from Alice to Bob. Whether or not the quantum protocol approximates to the science fiction ideal, however, it remains a very remarkable phenomenon from the information-theoretic point of view\footnote{Interestingly, it can be argued that quantum teleporation is perhaps not so far from the sci-fi ideal as one might initially think. \citet{vaidman} suggests that if all physical objects are made from elementary particles, then what is distinctive about them is their form (i.e. their particular state) rather than the matter from which they are made. Thus it seems one could argue that \textit{objects} really are teleported in the protocol.}. For consider what has been achieved. An unknown quantum state has been sent to Bob; and how else could this have been done? Only by Alice sending a quantum system \textit{in} the state $\ket{\chi}{}$ to Bob\footnote{Or by her sending Bob a system in a state explicitly related to $\ket{\chi}{}$ (\citet[cf.][]{park:1970}).}, for she cannot determine the state of the system and send a description of it instead. (Recall, it is impossible to determine an unknown state of an individual quantum system. See \citet{busch:observable} for a nice review.)

If, however, Alice did \textit{per impossibile} somehow learn the state and send a description to Bob, then systems encoding that description would have to be sent between them. In this case something that \textit{does} bear a relation to the identity of the state is transmitted from Alice to Bob, unlike in teleportation. Moreover, sending such a description would require a \textit{very great deal} of classical information, as in order to specify a general state of a two dimensional quantum system, two \textit{continuous} parameters need to be specified.

The picture we are left with, then, is that in teleportation there has been a transmission of something that is inaccessible at the classical level (often loosely described as a transmission of \textit{quantum} information); in the transmission this information has been in some sense disembodied; and finally, the transmission has been very efficient---requiring, apart from prior shared entanglement, the transfer of only two classical bits.

\subsection{Some information-theoretic aspects of teleportation}      
\subsubsection{Preamble}
The notion of information that is central to quantum information theory is that deriving from the seminal work of \citet{shannon} in communication theory. He introduced a measure of information $H(X)$ to characterise a source $X$ of messages which are produced from a fixed alphabet $\{x_{1},\ldots ,x_{n}\}$ whose elements occur with probability $p(x_{i})$. The Shannon information $H(X)$ measures in bits (classical two-state systems) the resources required to transmit all the messages that the source produces (Shannon's \textit{noiseless coding theorem} see Appendix~\ref{elements}). That is, it measures how much the messages from the source can be compressed. Shannon also introduced the \textit{mutual information}, $H(X:Y)$, which indicates how much information it is possible to transmit over a noisy channel (intuitively: how much can be inferred about the input to a channel, given the output obtained).

The Shannon information measure has many important applications in the quantum context\footnote{Surprisingly, perhaps, this point has occasioned some controversy (\citet{conceptualinadequacy}). See \citet{supposed} for a detailed discussion of the applicability of the Shannon information in the quantum context.}, for example, when considering the transmission of classical (Shannon) information over a channel consisting of quantum devices; but it is also possible to introduce an important and closely related concept---that of \textit{quantum} information properly so-called. This was done by Schumacher in the early 1990s\footnote{Published as \citet{qcoding}, \citet{schumacher:jozsa}.}.

Schumacher followed Shannon's lead: consider a device---a \textit{quantum source}---which, rather than outputting systems corresponding to elements of a classical alphabet, produces systems in particular quantum states $\rho_{x_{i}}$ with probabilities $p(x_{i})$. 
By reasoning analogous to Shannon's, Schumacher showed that the output of this source could be compressed by an amount measured by the von Neumann entropy of the source (the \textit{quantum} noiseless coding theorem, Appendix~\ref{elements}). We therefore have, analogously to the classical case, a notion of the amount of quantum information that the source produces: a measure of the minimum number of two-state quantum systems (\textit{qubits}) required to encode the output of the source.   

While it is important to distinguish these technical notions of information from the everyday notion of information linked to knowledge, language and meaning (\textit{pace} \citet{dretske:1981}, cf. \citet{thesis} Chpt.1) there is at least one interesting property held in common: in both the technical and everyday settings `information' functions as an abstract noun, that is, as a term which does not serve to denote a kind of entity having a location in space and time. Briefly: in the latter case, because `information' is a nominalization of the verb `to inform'; in the former, 
because an answer to the question `what is transmitted?' will refer to an abstract type rather than a concrete thing; and because what is measured is a property of a source (compressibility) or a channel (capacity), not an amount of some stuff present in a message\footnote{I will not seek to defend these views in more detail here. See \citet{thesis} Part 1 \textit{passim} and Section 1.2.3 esp. for further discussion.}.     

\subsubsection{Application to Teleportation}

There are two information-theoretic aspects of the teleporation protocol it is helpful to go into in somewhat more detail. 
The first concerns our reason for saying that a very large amount of information is required to specify the state that is teleported. 

As we have noted, in order to describe an arbitrary (pure) state of a two dimensional quantum system, it is necessary to specify two continuous parameters. A useful means of picturing this is via the Bloch sphere representation. The pure states of a two-state quantum system are in one-to-one correspondence with the points on the surface of the unit 3-sphere, and we may specify two real numbers (angles) to determine a point on the sphere. But why should doing this have associated with it an amount of information? If it is to do so we will need to imagine a classical information source that is selecting these pairs of angles with various probabilities; then a certain Shannon information may be ascribed to the process. Given a particular output of this information source, a quantum system is prepared in the state corresponding to the two angles selected. The quantum states prepared in this manner will then have associated with them a \textit{specification information}\footnote{Cf. \citet{qcoding}, \citet{supposed} Section 4.1.} given by the information of the source. Once a system has been prepared in some state in this way, it is presented to Alice, who may proceed to teleport the state to Bob.

Rather than the pairs of angles being selected from their full, continuous, range of possible values, the surface of the sphere might be coarse-grained evenly to give a finite number of choices. One might pick the angles specifying the mid-point, say, of each small element of surface area to provide the finite set of pairs of angles to choose between. Loosely speaking this coarse-graining corresponds to considering angles only to a certain degree of accuracy. As this accuracy is increased (the choices made more finely grained), the number of bits required to specify the choice increases without bound. If our information source is selecting states to an arbitrarily high accuracy then, the specification information is unboundedly large. (On the other hand, if the information source is only selecting between a small number of distinct states, then the specification information is correspondingly small. From now on we will assume that unless otherwise stated, the unknown states to be teleported are selected from a suitable coarse-graining of the whole range of possible angles.) It is essential to note, however, that even if the specification information associated with the state that has been teleported to Bob is exceedingly large, the majority of this information is not accessible to him. This leads on to the second point.

When one considers encoding classical information in quantum systems, it is necessary to distinguish between specification information and \textit{accessible information}\footnote{Cf. \citet{qcoding}.}. The specification information refers to the information of the classical source that selects sequences of quantum states, the accessible information to the maximum amount of information that is available following measurements on the systems prepared in these states. Because of the existence of non-orthogonal states, specification and accessible information may differ; and as it is impossible to distinguish  non-orthogonal states perfectly, the specification information associated with a string of systems selected from a source may be much greater than the accessible information. In teleportation, the systems are prepared near Alice before teleportation of their states to Bob. He may then perform various measurements to try and learn something. Call the information of the source selecting the states to be teleported by Alice $H(A)$; the mutual information $H(A:B)$ will determine the amount of classical information per system that Bob is able to extract by performing some measurement, $B$, following successful teleportation of the unknown state. The accessible information is given by the maximum over all decoding measurements of $H(A:B)$. A well known result due to \citet{holevo}---the Holevo bound---restricts the amount of information that Bob may acquire to a maximum of one bit of information per qubit, that is, to a maximum of one bit of information per successful run of the teleportation protocol.

So this gives us the sense in which the very large amount of information that may be associated with the unknown state being teleported to Bob is largely inaccessible to him. Note that the amount of information that Bob may acquire from the teleported state is less than the amount of classical information---two bits---that Alice had to send to him during the protocol. This fact is of the utmost importance, for if the Holevo bound did not guarantee this, and Bob were able to extract more than two bits of information from his system, then teleportation would give rise to paradox (when embedded in a relativistic theory) as superluminal signalling would be possible\footnote{The argument parallels the one given by \citet{teleportation} to the effect that two full classical bits are required in teleportation. In essence, if Bob were able to gain more than two bits of information in the protocol, then even if he were not to wait for Alice to send him the pair of bits each time and simply guessed their values instead, then some information would still get across.}.

So the Holevo bound ensures that teleportation is not paradoxical, but it also means that teleportation, when considered as a mode of ordinary \textit{classical} information transfer, is pretty inefficient, requiring two classical bits to be sent for every bit of information that Bob can extract at his end.

\section{The puzzles of teleportation}\label{puzzles}

Let us return to the picture of teleportation that was sketched earlier. An unknown quantum state is teleported from Alice to Bob with nothing that bears any relation to the identity of the state having travelled between them. The two classical bits sent are quite insufficient to specify the state teleported; and in any case, their values are independent of the parameters describing the unknown state. The unboundedly large specification information characterizing the state---information that is inaccessible at the classical level---has somehow been disembodied, and then reincarnated at Bob's location, as the quantum state first `disappears' from Alice's system and then `reappears' with Bob.

The conceptual puzzles that this process presents seem to cluster around two essential questions. First, how is \textit{so much} information transported? And second, most pressingly, just \textit{how} does the information get from Alice to Bob?

Perhaps the prevailing view on how these questions are to be answered is the one that has been expressed by \citet{jozsa:1998,jozsa:2003} and \citet{penrose:1998}. In their view, the classical bits used in the protocol evidently can't be carrying the information, for the reasons we have just rehearsed; therefore the entanglement shared between Alice and Bob must be providing the channel down which the information travels. They conclude that in teleportation, an indefinitely large, or even infinite amount of information travels backwards in time from Alice's measurement to the time at which the entangled pair was created, before propagating forward in time from that event to Bob's performance of his unitary operation and the attaining by his system of the correct state. Teleportation seems to reveal that entanglement has a remarkable capacity to provide a hitherto unsuspected type of information channel, which allows information to travel backwards in time; and a very great deal of it at that. Further, since it is a purely quantum link that is providing the channel, it must be purely \textit{quantum} information that flows down it. It seems that we have made the discovery that quantum information is a new \textit{type} of information with the striking, and non-classical, property that it may flow backwards in time.   

The position is summarized succinctly by Penrose:

\begin{quoting}
`How is it that the \textit{continuous} ``information" of the spin direction of the state that she [Alice] wishes to transmit...can be transmitted to Bob when she actually sends him only two bits of discrete information? The only other link between Alice and Bob is the quantum link that the entangled pair provides. In spacetime terms this link extends back into the past from Alice to the event at which the entangled pair was produced, and then it extends forward into the future to the event where Bob performs his [operation].

Only \textit{discrete} classical information passes from Alice to Bob, so the complex number ratio which determines the specific state being ``teleported" must be transmitted by the \textit{quantum} link. This link has a channel which ``proceeds into the past" from Alice to the source of the EPR pair, in addition to the remaining channel which we regard as ``proceeding into the future" in the normal way from the EPR source to Bob.There is no other physical connection.'
(\citet[p.1928]{penrose:1998})
\end{quoting}

But one might feel, with good reason, that this explanation of the nature of information flow in teleportation is simply too outlandish. This is the view of \citet{dh}, who conclude instead that with suitable analysis, the message sent from Alice to Bob can, after all, be seen to carry the information characterizing the unknown state. The information flows from Alice to Bob hidden away, unexpectedly, in Alice's message. This approach, and the question of what light it may shed on the notion of quantum information, has been considered in detail elsewhere (\citet{nifpaper}). Suffice it to say for present purposes that Deutsch and Hayden disagree with Jozsa and Penrose over the nature of quantum information and how it may flow in teleportation.

One might adopt yet a third, and perhaps more prosaic response to the puzzles that teleportation poses. This is to adopt the attitude of the \textit{conservative classical quantity surveyor}\footnote{A resolution along these lines, tied also to an ensemble view of the quantum state (\textit{vide infra}) has been suggested by \citet{jon1} and \citet{petermorgan}.\label{ccqs ensemble}}. According to this view, an amount of information cannot be said to have been transmitted to Bob unless it is accessible to him. But of course, as we noted above, the specification information associated with the state teleported to Bob is \textit{not} accessible to him: he cannot determine the identity of the unknown state. On this view, then, the information associated with selecting some unknown state $\ket{\chi}{}$ will not have been transmitted to Bob until an entire ensemble of systems in the state $\ket{\chi}{}$ has been teleported to him, for it is only then that he may determine the identity of the state\footnote{Note that we will need to adjust our scenario slightly to incorporate this view. In our initial set-up, the source $A$ selected a sequence of states which were then teleported one by one to Bob. Now we imagine instead that following some particular output of $A$, an entire ensemble of systems is prepared in the pure state associated with that output; then this ensemble of systems --- all in the same unknown pure state --- is teleported. This adjustment is required because in our initial set-up for the teleportation procedure, the only way in which an ensemble of systems all in the same state could be teleported to Bob would be by setting the information of the source $A$ to zero, with the tiresomely paradoxical result that Bob could now determine the state all right, but would gain no information by doing so.\label{noinfo}}. To teleport a whole ensemble of systems, though, Alice will need to send Bob an infinite number of classical bits; and now there isn't a significant disparity between the amount of information that has been explicitly sent by Alice and the amount that Bob ends up with. One needs to send a very large number of classical bits to have transmitted by teleportation the very large amount of information associated with selecting the unknown state. 

This approach does not seem to solve all our problems, however. Someone sympathetic to the line of thought espoused by Jozsa and Penrose can point out in reply that there still remains a mystery about \textit{how} the information characterizing the unknown state got from Alice to Bob---the bits sent between them, recall, have no dependence on the identity of the unknown state. So while the approach of the conservative classical quantity surveyor may mitigate our worry to some extent over the first question, it does not seem to help with the second.

\section{Resolving (dissolving) the problem}\label{dissolving}

Dwelling on the question of how the information characterizing the unknown state is transmitted from Alice to Bob has given rise to some conundrums. Should we side with Jozsa and Penrose and admit that quantum information may flow backwards in time down a channel constituted by shared entanglement? Or perhaps with Deutsch and Hayden, and agree that information should flow in a less outlandish fashion, but that quantum information may be squirrelled away in seemingly classical bits? Counting conservatively the amounts of information available after teleportation may make us less anxious about the load carried in a single run of the protocol, but the question still remains: how did the information, in the end, get to Bob? Should we just conclude that it is transported nonlocally in some way? But what might that mean?

If the question `How does the information get from Alice to Bob?' is causing us these difficulties, however, perhaps it might pay to look at the question itself rather more closely. In particular, let's focus on the crucial phrase `the information'.

Our troubles arise when we take this phrase to be referring to a particular, to some sort of substance or entity whose behaviour in teleportation it is our task to describe. The assumption common to the approaches of Deutsch and Hayden on the one hand, and Jozsa and Penrose on the other, is that we need to provide a story about how some \textit{thing} denoted by `the information' travels from Alice to Bob. Moreover, it is assumed that this supposed thing should be shown to take a spatio-temporally continuous path.

But notice that `information' is an abstract noun. This means that `the information' certainly does \textit{not} refer to a substance or to an entity. The shared assumption is thus a mistaken one, and is based on the error of hypostatizing an abstract noun. 
If `the information' doesn't introduce a particular, then the question `How does the information get from Alice to Bob?' cannot be a request for a description of how some thing travels. It follows that the locus of our confusion is dissolved.

But if it is a mistake to take `How does the information get from Alice to Bob?' as a question about how some thing is transmitted, then what is its legitimate meaning, if any? It seems that the only legitimate use that can remain for this question is as a flowery way of asking: what are the physical processes involved in the transmission? Now \textit{this} question is a perfectly straightforward one, even if, as we shall see (Section~\ref{interpretations}), the answer one actually gives will depend on the interpretation of quantum mechanics one adopts. But there is no longer a \textit{conceptual} puzzle over teleportation. Once it is recognised that `information' is an abstract noun, then it is clear that there is no further question to be answered regarding how information is transmitted in teleportation that goes beyond providing a description of the physical processes involved in achieving the aim of the protocol. That is all that `How is the information transmitted?' can intelligibly mean; for there is not a question of information being a substance or entity that is transported, nor of `the information' being a referring term. Thus, one does not face a double task consisting of a) describing the physical processes \textit{by which} information is transmitted, followed by b) tracing the path of a ghostly particular, information. There is only task (a). 

The point should not be misunderstood:  I am not claiming that there is no such thing as the transmission of information,
but simply that one should not understand the transmission of information on the model of transporting potatoes, or butter, say, or piping water\footnote{Note that we do sometimes talk of a flow of information; and we do say of many physical quantities that are not entities or substances --- for example, energy, heat --- that they flow. But there is no analogy between the two cases, for what this latter description means is that the quantities in question obey a local conservation equation. It is not clear that it is at all intelligible to suggest that information should obey a local conservation equation. Certainly, the concept of quantity of information that is provided by the Shannon theory does not give us a concept of a quantity it makes sense to suggest might obey such an equation. (On this, see Section~\ref{study concluding} below.)}.

\subsection{The simulation fallacy}\label{simulation fallacy}

Whilst paying due attention to the status of `information' as an abstract noun provides the primary resolution of the problems that teleportation can sometimes seem to present us with, there is a secondary possible source of confusion that should be noted. This is what may be termed the \textit{simulation fallacy}.

Imagine that there is some physical process $\cal{P}$ (for example, some quantum-mechanical process) that would require a certain amount of communication or computational resources to be simulated classically. Call the classical simulation using these resources $\cal{S}$. The simulation fallacy is to assume that because it requires these classical resources to simulate $\cal{P}$ using $\cal{S}$, there are processes going on when $\cal{P}$ occurs that are physically equivalent to (are instantiations of) the processes that are involved in the simulation $\cal{S}$ itself (although these processes may be being instantiated using different properties in $\cal{P}$). In particular, when $\cal{P}$ is going on, the thought is that there must be, at some level, physical processes involved in $\cal{P}$ which correspond concretely to the evolution of the classical resources in the simulation $\cal{S}$.
The fallacy is to read off features of the simulation as real features of the thing simulated\footnote{Note that it will not always be fallacious to take features of a simulation to correspond to features of the simulated --- if the features in question are explicitly \textit{analogues} of features of the system or process being simulated. One should thus distinguish between i) simulations that involve analogues and ii) functional `black-box', or input-output simulations.}.

A familiar example of the simulation fallacy is provided by Deutsch's argument that Shor's factoring algorithm supports an Everettian view of quantum mechanics (\citet[p.217]{FoR}). The argument is that if factoring very large numbers would require greater computational resources than are contained in the visible universe, then how could such a process be possible unless one admits the existence of a very large number of (superposed) computations in Everettian parallel universes? A computation that would require a very large amount of resources if it were to be performed classically is explained \textit{as} a process that consists of a very large number of classical computations. But of course, considered as an argument, this is fallacious. The fact that a very large amount of classical computation might be required to produce the same result as a quantum computation does not entail that the quantum computation consists of a large number of parallel classical computations\footnote{For further discussion of Deutsch's conception of quantum `parallel processing', see \citet{steane} and \citet{horsman}.}.   

The simulation fallacy is also evident in the common claim that Bell's theorem shows us that quantum mechanics is nonlocal, or the claim that the experimental violation of Bell inequalities means that the world must be nonlocal. Of course, what is in fact shown by these well-known results is that no local hidden variable model can simulate the predictions of quantum mechanics, nor provide a model for the experimentally observed correlations. But these facts about simulation don't lead directly to facts about the simulated: the fact that any adequate hidden variable model must be nonlocal does not show that quantum mechanics is nonlocal (this, of course, is an interpretation dependent property), nor show the world to be nonlocal.

While the question of what classical resources would be required to simulate a given quantum process is an indispensible guide in the search for interesting quantum information protocols and is vitally important for that reason, the simulation fallacy indicates that it is by no means a sure guide to ontology. 

With regard to teleportation, it is important to recognise the simulation fallacy in order to assuage any worries that might remain over the question `How does so much information get from Alice to Bob?', and to undermine further the thought that teleportation must be understood as a flow of information. 

For the fact that it would take a very large number of classical bits to transmit the identity of an unknown state from Alice to Bob does not entail that
in teleportation there is a real corresponding transmission of information, some physical process going on that instantiates, albeit in a different medium, the transport of this large amount of information\footnote{Nor, for example, does the fact that there are protocols in which the state of a qubit can be substituted for an arbitrarily large amount of classical information (\citet{lucien:substituting}) imply that this large amount of information is really there in the qubit. }. (Note that the flow of the hypostatized `quantum information' of Jozsa and Penrose plays precisely this r\^ole: the analogue, in a different medium, of the transport of the large amount of classical information.) Equivalence from the point of view of information processing does not imply physical equivalence. 

Awareness of the simulation fallacy is particularly relevant when we consider the approach of the conservative classical quantity surveyor. Recall that the point of this approach is to deny that a large amount of information can be said to have been transported to Bob in teleportation until that information is actually available to him. However, it might be objected to this that after a single run of the teleportation protocol, the information characterizing the state is certainly present at Bob's location, even if inaccessible to him, as a system \textit{in} the unknown state is present\footnote{It is for this reason that it is natural to marry conservative classical quantity surveying with an ensemble view of the quantum state (see footnote~\ref{ccqs ensemble}), for then this objection would not go through --- when the two positions are conjoined, not only is the information characterizing the state not available until the whole ensemble is teleported, but neither has the \textit{state} been teleported until the whole ensemble has been teleported to Bob.}.

This contention would seem to rest on an argument of the following form:
The only way the unknown state can appear at Bob's location is if the information characterizing the state has actually been transported to Bob, hence on appearance of the state, the specification information associated with the state has indeed been transported to Bob's location. (Crudely, if a system in the given state is present, then the information is present, as it takes this information to specify the state.) But such an argument needs to be treated with care, for the main premise appears to rest on the simulation fallacy. Just because it would take a large amount of information to specify a state doesn't mean that we should conclude that this amount of information has been physically transported in teleportation when Bob's system acquires the state.

In any event, the simplest way to remain clear on whether or not, or in what way, information can be said to be present at Bob's location following a single run of the teleportation protocol is to respect the distinction between the specification information associated with a system and the amount of information that may be said to be encoded or contained in the system. Once Bob's system has acquired the state $\ket{\chi}{}$ teleported by Alice, then his system has associated with it the same specification information, $H(A)$: \textit{if} one were now \textit{asked} to specify the state of Bob's system, then this number of bits would be required, on average. This quantity of information is not encoded or contained in the system however. The mutual information $H(A:B)$ and the accessible information provide the relevant measures of how much information Bob's system can be said to contain, for they govern the amount that may be decoded. But of course, as `information' is an abstract noun, containing information is not containing some \textit{thing}, however aethereal.

\section{The teleportation process under different interpretations}\label{interpretations}

By reflecting on the logico-grammatical status of the term `information' we have been able to replace the (needlessly) conceptually puzzling question of how the information gets from Alice to Bob in teleportation, with the simple, genuine, question of what the physical processes involved in teleportation are. While this may not, perhaps, be quite enough to still all the controversy that trying to understand teleportation has evoked, the controversy is now of a very familiar kind: it concerns what interpretation of quantum mechanics one adopts. For the detailed story one tells about the physical processes involved in teleporation will of course depend upon one's interpretive stance. Two questions in particular will find different answers under different interpretations: first, is nonlocality involved in teleportation? and second, has anything interesting happened before Alice's classical bits are sent to Bob and he performs the correct unitary operation? 

We will now see how some of these differences play out in the following familiar interpretations (the list of approaches considered is by no means exhaustive).   

\subsection{Collapse interpretations: Dirac/von Neumann, GRW}

The natural place to begin is with the orthodox approach of \citet{dirac} and \citet{vN} in which there is a genuine process of collapse on measurement\footnote{One of the defining features of what I here term `orthodoxy' is the adoption of the standard eigenstate-eigenvalue link for the ascription of definite values to quantum systems. See e.g. \citet{bub:1997}.}. (The vagueness over where, when, why and how this collapse takes place might be alleviated along lines suggested by \citet{grw}, perhaps.) If one has a genuine process of collapse then as noted long ago by \citet*{EPR}\footnote{See \citet{erpart1} for a recent discussion.}, one has action-at-a-distance. In the presence of entanglement, a measurement on one system can result in a real change to the possessed properties of another system, even when the two systems are widely separated. (Although, as is well known, these changes do not allow one to send signals superluminally---this is known as the \textit{no-signalling theorem}\footnote{An early version of the no-signalling theorem, specialised to the case of spin 1/2 EPR-type experiments appears in \citet{bohm}. Later, more general versions are given by \citet{tausk,eberhard,grw:no-signalling}. See also \citet{shimony}, \citet{redhead:no-signalling}.}.) 

In teleportation, then, under a collapse interpretation, the effect of Alice's Bell-basis measurement will be to prepare Bob's system, at a distance, in one of four pure states which depend on the unknown state $\ket{\chi}{}$, by using the nonlocal effect of collapse. It then only remains for Alice to send her two bits to Bob to tell him which (type of) state he now has in his possession. Under this interpretation, teleportation explicitly involves nonlocality, or action-at-a-distance; and it is precisely because of the nonlocal effect of collapse, preparing Bob's system in a state that differs in one of only four ways from $\ket{\chi}{}$, that a mere two classical bits need be sent by Alice in order for Bob's system to acquire a state parameterised by two continuous values.

It is enlightening to compare the effect of collapse in this scenario to that of a rigid rod held by two parties. Imagine that Alice wanted to let Bob know the value of a parameter that could take on values in the interval $[0,1]$. If they were each holding one end of a long rigid rod, then Alice could let Bob know the value she has in mind simply by moving her end of the rod along in Bob's direction by a suitable distance. Bob, seeing how far his end of the rod moves, may infer the value Alice is thinking of\footnote{Of course, in a relativistic setting, rigid bodies would not be permissible, although they are in non-relativistic quantum mechanics. This does not in any case affect the point of the analogy.}. There is no mystery here about how the value of the continuous parameter is transmitted from Alice to Bob. Alice, by moving her end of the rod, moves Bob's by a corresponding amount. In teleportation, the effect of collapse is somewhat analogous: Bob's system is prepared, by the nonlocal effect of collapse, in a state that depends on the two continuous parameters characterizing $\ket{\chi}{}$. As we have said, collapse allows a real change in the physical properties that a distant system possesses, if there was prior entanglement. Compare: pushing one end of a rigid rod axially leads to a change in the position of the far end. The nonlocal effect of collapse, which is here understood as a real physical process, is providing the main physical mechanism behind teleportation; and recall that once the physical mechanisms have been described (I have argued) there is no further question to be asked about how information is transmitted in the protocol.       

In a collapse interpretation, teleportation thus involves nonlocality, in the sense of action-at-a-distance, crucially. Also, something interesting certainly has happened once Alice performs her measurement and before she sends the two classical bits to Bob. There has been a real change in the physical properties of Bob's system, as it acquires one of four pure states. (Although note that at this stage the probability distributions for measurements on Bob's system will nonetheless not display any dependence on the parameters characterizing $\ket{\chi}{}$, in virtue of the no-signalling theorem. It is only once the bits from Alice have arrived and Bob has performed the correct operation that measurements on his system will display a dependence on the parameters $\alpha$ and $\beta$.)

\subsection{No collapse and no extra values: Everett}

It is possible to retain the idea that the wavefunction provides a complete description of reality while rejecting the notion of collapse; this way lies the Everrett interpretation (\citet{everett})\footnote{It should be noted that there have been a number of different attempts to develop Everett's original ideas into a full-blown interpretation of quantum theory. The most satisfactory of these would appear to be an approach on the lines of Saunders and Wallace (\citet{simon1,simon2,simon3,simonrelativism,wallace:worlds,wallace:structure}) which resolves the preferred basis problem and has made considerable progress on the question of the meaning of probability in Everett (on this, see in particular \citet{deutsch:decision,wallace:rationality}).}. The characteristic feature of the Everett interpretation is that the dynamics is always unitary; and no extra values are added to the description provided by the wavefunction in order to account for definite measurement outcomes. Instead, measurements are simply unitary interactions which have been chosen so as to correlate states of the system being measured to states of a measuring apparatus. Obtaining a definite value on measurement is then understood as the measured system coming to have a definite state (eigenstate of the measured observable) \textit{relative to} the indicator states of the measuring apparatus and ultimately, relative to an observer\footnote{This is the case for ideal first-kind (non-disturbing) measurements. The situation becomes more complicated when we consider the more physically realistic case of measurements which are not of the first kind; in some cases, for example, the object system may even be destroyed in the process of measurement. What is important for a measurement to have taken place is that measuring apparatus and object system were coupled together in such a way that if the object system had been in an eigenstate of the observable being measured prior to measurement, then the subsequent state of the measuring apparatus would allow us to infer what that eigenstate was. In this more general framework the importance is not so much that the object system is left in a eigenstate of the observable relative to the indicator state of the measuring apparatus, but that we have definite indicator states relative to macroscopic observables.}. 
A treatment of teleportation in the Everettian context was given by \citet{vaidman}. \citet{braunstein:irreversible} provides a detailed discussion of the teleportation protocol within unitary quantum mechanics without collapse.

With teleportation in an Everettian setting, and unlike teleportation under the orthodox account, it is clear that there will be no action-at-distance in virtue of collapse when Alice performs her measurement, for the simple reason that there is no process of collapse. Instead, the result of Alice's measurement will be that Bob's system comes to have definite relative states related to the unknown state $\ket{\chi}{}$, with respect to the indicator states of the systems recording the outcome of Alice's measurement (see Appendix~\ref{nocollapse appendix}). (It is argued in \citet{erpart1} and \citet{nifpaper} that this does not amount to a new form of nonlocality.) Note, though, that at this stage of the protocol, the \textit{reduced} state of every system involved will now be maximally mixed\footnote{This would not in general be the case if the initial entangled state were not maximally entangled, or if Alice's measurement were not an ideal measurement; with these eventualities, the teleportation would be imperfect (fidelity less than 1).}. As \citet{braunstein:irreversible} notes, this feature corresponds to the `disembodiment' of the information characterizing the unknown state in the orthodox account of teleportation: following Alice's measurement, all the systems involved in the protocol will have become fully entangled. Dependence on the parameters characterizing the unknown state will only be observable with a suitable \textit{global} measurement, not for any local measurements. In particular, one can consider the correlations that now exist between the systems recording the outcome of Alice's measurement and Bob's system. Certain of the joint (and irreducible) properties of these spatially separated systems will depend on the identity of the unknown state. In this sense, the information characterizing $\ket{\chi}{}$ might now be said to be `in the correlations' between these systems. (This is the terminology Braunstein adopts.) 

Once Bob has been sent the systems recording the outcome of Alice's measurement, however, he is able to disentangle his system from the other systems involved in the protocol. Its state will now factorise from the joint state of the other systems; and will in fact be the pure state $\ket{\chi}{}$. Dependence on the parameters $\alpha$ and $\beta$ \textit{will} finally be observable for local measurements once more, but this time, only at Bob's location. 

In collapse versions of quantum mechanics, the nonlocal effect of collapse was the main physical mechanism underlying teleportation. In the no-collapse Everettian setting, the fundamental mechanism is provided by the fact that in the presence of entanglement, local unitary operations---in this case, Alice's measurement---can have a non-trivial effect on the global state of the joint system.

So, has anything significant happened at Bob's location before Alice sends him the result of her measurement and he performs his conditional unitary operation? Well, arguably not: nothing has happened other than all of the systems involved in the protocol having become entangled, as a result of the various local unitary operations.

\subsection{No collapse, but extra values: Bohm}

The Bohm theory account provides us with an interesting intermediary view of teleportation, in which there is no collapse of the wavefunction, but nonlocality plays an interesting r\^ole. We shall follow the analysis of \citet{maroney:hiley}.

The Bohm theory (\citet{bohm:1952}) is a nonlocal, contextual, deterministic hidden variable theory, in which the wavefunction $\Psi(\mathbf{x}_{1},\mathbf{x}_{2}\ldots\mathbf{x}_{n},t)$ of an $n$-body system evolves unitarily according to the Schr\"odinger dynamics, but is supplemented with definite values for the positions $\mathbf{x}_{1}(t),\mathbf{x}_{2}(t)\ldots\mathbf{x}_{n}(t)$ of the particles. Momenta are also defined according to $\mathbf{p}_{i}=\nabla_{i}S$, where $S$ is the phase of $\Psi$, hence a definite trajectory may be associated with a system, where this trajectory will depend on the many-body wavefunction (and thus, in general, on the positions and behaviour of all the other systems, however far away). If the initial probability distribution for particle positions is assumed to be given by $|\Psi|^{2}$, then the same predictions for measurement outcomes will be made as in ordinary quantum mechanics. For detailed presentations of the Bohm theory, see \citet{bohmhiley} and \citet{holland:1995}.

The guiding effect of the wavefunction on the particle positions may also be understood in terms of a new \textit{quantum potential} that acts on particles in addition to the familiar classical potentials. The quantum potential is given by 
\[Q(\mathbf{x}_{1},\mathbf{x}_{2},\ldots,\mathbf{x}_{n})=-\hbar^{2}\sum_{i=1}^{n}\frac{\nabla^{2}_{i}R}{2m_{i}R},\]
where $R$ is the amplitude of $\Psi$ and $m_{i}$ is the mass of the $i$-th particle. Among the ways in which this quantity differs from a classical potential is that it will in general give rise to a nonlocal dynamics (that is, in the presence of entanglement, the force on a given system will depend on the instantaneous positions of the other particles, no matter how far away); and it may be large even when the amplitude from which it is derived is small. \citet[\S 3.2]{bohmhiley} suggest that the quantum potential should be understood as an `information potential' rather than a mechanical potential, as a way of accounting for its peculiar properties.

The determinate values for position in the Bohm theory are usually understood as providing the definite outcomes of measurement\footnote{Note, though, that measurement may not usually be understood as revealing pre-existing values in the Bohm theory. Interestingly, \citet{bw:2005} have recently argued that these definite position values may not be so helpful in solving the measurement problem as is often supposed.} that would appear to be lacking in a no-collapse version of quantum mechanics, in the absence of an Everett-style relativization. Following a measurement interaction, the wavefunction of the joint object-system and apparatus will have separated out (in the ideal case) into a superposition of non-overlapping wavepackets (on configuration space) corresponding to the different possible outcomes of measurement.
The determinate values for the positions of the object-system and apparatus pointer variable will pick out a point in configuration space; and the outcome that is observed, or is made definite, is the one corresponding to the wavepacket whose support contains this point. The wavefunction for the total system remains as a superposition of all of the non-overlapping waverpackets, however. \citet{bohmhiley} introduce the notions of \textit{active}, \textit{passive} and \textit{inactive} information to describe this feature of the theory. If $\Psi$ may be written as a superposition of non-overlapping wavepackets, then they suggest that the definite configuration point of the total system picks out one of these wavepackets (the one whose support contains the point) as active. The evolution of the point is determined solely by the wavepacket containing it; and in keeping with their conception of $Q$ as an information potential, the information associated with this wavepacket is said to be active. The information associated with the other wavepackets is termed either `passive', or `inactive'. `Passive', if the different wavepackets may in the future be made to overlap and interfere, `inactive' if such interference would be a practical impossibility (as for example, if environmental decoherence has occurred in a measurement-type situation --- this corresponds to the case of `effective collapse' of the wavefunction). 

In their discussion of the teleportation protocol, Maroney and Hiley adopt the approach in which a definite spin vector is also associated with each spin 1/2 particle, in addition to its definite position.
The idea is that with each system is associated an orthogonal set of axes (body axes) whose orientation is specified by a real three dimensional spin vector, $\mathbf{s}$, along with an angle of rotation about this vector; where these quantities are determined by the wavefunction\footnote{This is the approach to spin of \citet{bst:1955}. For a systematic presentation see \citet[\S 10.2-10.3]{bohmhiley} or \citet[Chpt. 9]{holland:1995}. Other approaches to spin are possible, e.g., \citet[\S 10.4-10.5]{bohmhiley}, \citet[Chpt. 10]{holland:1995}, or the `minimalism' of \citet{bell:1966,bell:1981}, in which no spin values are added.}.  

The analysis of teleportation then proceeds much as in the Everett interpretation, save that we may also consider the evolution of the determinate spin vectors associated with the various systems. Initially, system 1 in the unknown state $\ket{\chi}{}$ will have some definite spin vector that depends on $\alpha$ and $\beta$, $\mathbf{s}(\alpha,\beta)$, while it turns out that if Alice and Bob share a singlet state, the spin vectors for their two systems will be zero (\citet[\S10.6]{bohmhiley}). Now Alice performs her Bell-basis measurement. As in the Everettian picture, the effect of measurement is to entangle the systems being measured with systems recording the outcome of the measurement. But this is not the only effect, in the Bohm theory. The total wavefunction is now a superposition of four terms corresponding to the four possible outcomes of Alice's measurement; and one of these four terms will be picked out by the definite position value of the measuring apparatus pointer variable. For each of these four terms taken individually, Bob's system will be in a definite state related to the state $\ket{\chi}{}$, thus with each will be associated a definite spin vector $\mathbf{s}^{j}(\alpha,\beta)$, $j=1,\ldots,4$, pointing in some direction. When one of the four terms is picked out as active, and the others rendered passive (or inactive), following Alice's measurement, the spin vector for Bob's system will change instantaneously from zero to one of the four $\mathbf{s}^{j}(\alpha,\beta)$ (\citet{maroney:hiley}).        

Thus in the Bohm theory, teleportation certainly involves nonlocality; and moreover, something very interesting does happen as soon as Alice has made her measurement. Bob's system acquires a definite spin vector that depends on the parameters characterizing the unknown state, as a result of a nonlocal quantum torque (\citet{maroney:hiley}). Furthermore, there is a one in four chance that this spin vector will be the same as the original $\mathbf{s}(\alpha,\beta)$; and all this while the total state of the system remains uncollapsed, with all the particles entangled. 

Finally, as we have seen before, once Alice sends Bob systems recording the outcome of her measurement, he may perform the conditional unitary operation necessary to disentangle his system from the others, and leave his system in the state $\ket{\chi}{}$. The spin vector of his system will now be $\mathbf{s}(\alpha,\beta)$ with certainty.

\subsubsection{A note on active information}\label{active information}

The conclusion of \citet{maroney:hiley} and \citet{hiley:1999} is that according to the Bohm theory, what is transferred from Alice's region to Bob's region in the teleportation protocol is the active information that is contained in the quantum state of the initial system. However questions may be raised about how apposite this description is.

Let us label the pointer degree of freedom of the measuring apparatus by $x_{0}$.
At the beginning of the teleportation protocol, the state of system 1 factorises from the entangled joint state of 2 and 3; and the state of the measuring apparatus will also factorise. Accordingly, the quantum potential will be given by a sum of separate terms:
\begin{equation}\label{qpotential before}
Q(\mathbf{x}_{1},\mathbf{x}_{2},\mathbf{x}_{3},x_{0}) = Q(\mathbf{x}_{1},\alpha,\beta) + Q(\mathbf{x}_{2},\mathbf{x}_{3}) + Q(x_{0}),
\end{equation}
where it has been noted that the first term, the one that will determine the motion of system 1, depends on the parameters characterizing the unknown state\footnote{The component of the force on the $i$-th system due to the quantum potential is given by $m_{i}\ddot{\mathbf{x}}_{i}=-\nabla_{i}Q$ (\citet[cf.][\S 7.1.2]{holland:1995}); therefore, only terms in the sum which depend on $\mathbf{x}_{i}$ will contribute to the motion of the $i$-th system.}.

Once Alice performs her Bell basis measurement, however, all the systems become entangled; and the potential will be of the form:
\begin{equation}\label{qpotential middle}
Q(\mathbf{x}_{1},\mathbf{x}_{2},\mathbf{x}_{3},x_{0}) = Q(\mathbf{x}_{1},\mathbf{x}_{2},x_{0}) + Q(\mathbf{x}_{3},x_{0},\alpha,\beta)
\end{equation}
The part of the quantum potential that will affect system 3 now depends on $\alpha$ and $\beta$.

Finally, at the end of the protocol, systems 1, 2 and the measuring apparatus are left entangled; and system 3, in the pure state $\ket{\chi}{3},$ factorises. The quantum potential then takes the form:
\begin{equation}\label{qpotential end}
Q(\mathbf{x}_{1},\mathbf{x}_{2},\mathbf{x}_{3},x_{0}) = Q(\mathbf{x}_{1},\mathbf{x}_{2},x_{0}) + Q(\mathbf{x}_{3},\alpha,\beta)
\end{equation}

Maroney and Hiley say:
\begin{quoting}
`What we see clearly emerging here is that it is active information that has been transferred from particle 1 to particle 3 and that this transfer has been mediated by the nonlocal quantum potential.'(\citet[p.1413]{maroney:hiley})

\noindent `...it is the objective active information contained in the wavefunction that is transferred from particle 1 to particle 3.' (\citet[p.1414]{maroney:hiley})
\end{quoting}

Note that the part of the potential that is active on system 3 will already have acquired a dependence on $\alpha$ and $\beta$ before the end of the protocol; that is, as soon as Alice has performed her measurement. So if active information depending on these parameters is transferred at all, it will have been transferred before the end of the protocol. However it is not until Alice has sent her message to Bob and he performs his conditional operation that the term  $Q(\mathbf{x}_{3},\alpha,\beta)$ in eqn.~(\ref{qpotential end}) will take the same form as the initial $Q(\mathbf{x}_{1},\alpha,\beta)$.

The difficulties for the stated conclusion arise when we consider more closely what is meant by `active information'. In \citet{maroney:hiley,hiley:1999}, the connection is made with a different sense of the word `information' than the ones we have considered so far. This is a sense that derives from the verb `inform' under its branch I and II senses (Oxford English Dictionary), \textit{viz.} to give form to, or, to give formative principle to (this latter, a Scholastic Latin offshoot).  

Thus `information' as it appears in `active information' and company, means the action of giving form to\footnote{Cf. OED `information', sense 7.}. `The information of $x$' (read: The \textit{in}-formation of $x$) means the action of giving form to $x$.

Now, while we may understand what is meant by $Q$ being said to be an information potential---it is a potential that gives form to something, presumably the possible trajectories associated with particles (although note that the distinction with mechanical potentials is now blurred, as these give form to the possible trajectories too)---and may understand the term `active' as picking out the part of the quantum potential that is shaping the actual trajectory in configuration space of the total system, it does not make sense to say that active information is transferred in teleportation. Because `information' here refers to a particular action---the giving of a form to something---and an action is not a \textit{thing} that can be moved\footnote{On some accounts, an action is the bringing about of some event or state of affairs by an agent (\citet{hyman:alvarez}); on others, an action is an event (\citet{davidson:actionsandevents}). On no account is an action something which can intelligibly be said to be moved about.}. The same \textit{type} of action may be taking place at two different places, or at two different times, but an action may not be moved from $A$ to $B$.   

Thus with `active information' understood in the advertised way,
all that can be said is that an action of the same \textit{type} is being performed (by the quantum potential) on system 3 at the end of the teleportation protocol as was being performed on system 1 at the beginning, not that something has been transferred between the two. We may not, then, understand `transfer' literally.  When all is said and done, it is perhaps clearer simply to adopt the standard description and say that the quantum \textit{state} of particle 1 has been `transferred' in teleportation; that is (as a quantum state is a mathematical object and therefore cannot literally be moved about either), that system 3 has been made to acquire (is left in) the unknown state $\ket{\chi}{}$.

To sum up: it perhaps looked as if the Bohmian notion of active information might provide us with a sense of what is transported in teleportation if we insist that \textit{information}, `the information in the wavefunction', is, in a literal sense, transported. But this proves not to be the case.

\subsection{Ensemble and statistical viewpoints}

So far, in all the interpretations we have considered, the quantum state may describe individual systems. Let us close this section by looking briefly at approaches in which the state is taken only to describe \textit{ensembles} of systems.

We may broadly distinguish two such approaches. The first I will term an \textit{ensemble} viewpoint. In this approach, the state is taken to represent a real physical property, but only of an ensemble. Following a measurement, the ensemble must be left in a \textit{proper} mixture\footnote{See \citet{d'Espagnat} for this terminology, also \citet{impsep}.}, in order for there to be definite outcomes, i.e., the ensemble is left in an appropriate mixture of sub-ensembles, each described by a pure state (eigenstate of the measured observable). Thus there will be a real process of collapse, but only at the level of the ensemble, not for individual systems (which are not being described by a quantum state, if at all). 

The second approach I call a \textit{statistical} interpretation. (This is the interpretation that would be adopted by instrumentalists, for example.) On this view, the quantum formalism merely describes the probabilities for measurement outcomes for ensembles, there is no description of individual systems and collapse does not correspond to any real physical process. 

On both these approaches, as the state is only associated with an ensemble, it is not until an entire ensemble has been teleported to Bob (that is, Alice has run the teleportation protocol on every member of an ensemble in the unknown state $\ket{\chi}{}$) that he acquires something in the state $\ket{\chi}{}$. An ensemble or statistical viewpoint thus makes a natural partner to conservative classical quantity surveying in teleportation.  

Under the statistical interpretation, there is clearly no nonlocality involved in teleportation, as there is no real process of collapse; and nothing of any interest has happened before the required classical bits are sent to Bob. (The no-signalling theorem entails that Alice's measurement won't affect the probability distributions for distant measurements.) The end result of the completed teleportation process is that Bob's ensemble is ascribed the state $\ket{\chi}{}$; where this merely means that the statistics one will expect for measurements on Bob's ensemble are now the same as those one would have expected for measurements on the initial ensemble presented to Alice.   
 
The ensemble viewpoint presents a rather different picture, as it does involve a real process of collapse, even if only at the ensemble level. Let us suppose that Alice has performed the Bell basis measurement on her ensembles, but has not yet sent the ensemble of classical bits to Bob. The effect of this measurement will have been to leave Bob's ensemble in a proper mixture composed of sub-ensembles in the four possible states a fixed rotation away from $\ket{\chi}{}$. Thus there has been a nonlocal effect: that of preparing what was an improper mixture into a particular proper mixture, whose components depend on the parameters characterizing the unknown state. The use of the flock of classical bits that Alice sends to Bob is to allow him to separate out the ensemble he now has into four distinct sub-ensembles, on each of which he performs the relevant unitary operation, ending up with all four being described by the state $\ket{\chi}{}$.

\section{Concluding remarks}\label{study concluding}

The aim of this paper has been to show how substantial conceptual difficulties can arise if one neglects the fact that `information' is an abstract noun. This oversight seems to lie at the root of much confusion over the process of teleportation; and this gives us very good reason to pay attention to the logical status of the term. A few closing remarks should be made.

Schematically, a central part of the argument has been of the following form:   

Puzzles arise when we feel the need to tell a story about how something travels from Alice to Bob in teleportation. 
In particular, it might be felt that this something needs to travel in a spatio-temporally continuous fashion; and one might accordingly feel pushed towards adopting something like the Jozsa/Penrose view.

But if `the information' doesn't pick out a particular, then there is no thing to take a path, continuous or not, therefore the problem is not a genuine one, but an illusion.

We can imagine a number of objections. A very simple one might take the following form: You have said that information is not a particular or thing, therefore it does not make sense to inquire how \textit{it} flows (but only inquire about the means by which it is transmitted). But don't we have a theory that quantifies information (\textit{viz.} communication theory); and if we can say how much of something there is, isn't that enough to say that we have a thing, or a quantity that can be located?

This objection is dealt with quickly. Note that this form of argument will not work in general---one can say how much a picture might be worth in pounds and pence, for example, but this is not quantifying an amount of stuff, nor describing a quantity with a location---and it does not work in this particular case either. The Shannon information doesn't quantify an amount of stuff that is present in a message, say, nor the amount of a certain quantity that is present at some spatial location. The Shannon information $H(X)$ describes a specific property of a \textit{source} (not a message), namely, the amount of channel resources that would be required to transmit the messages the source produces. This is evidently not to quantify an amount of stuff, nor to characterize a quantity that has a spatial location. (The source certainly has a spatial location, but its information does not.) Or consider the mutual information. Loosely speaking, this quantity tells us about the amount we may be able to infer about some event or state of affairs from the obtaining of another event or state of affairs. But how much we may infer is not a quantity it makes sense to ascribe a spatial location to.

Another objection might be as follows: You have suggested that it is a mistake to hypostatize information, to talk of it as a thing that moves about. How is this to be reconciled with some of the ways we often talk about information in physics, especially the example in relativity, where the most natural way of stating an important constraint is to say that relativity rules out the propagation of \textit{information} faster than the speed of light?

The response is that one can admit this mode of talking without it entailing a hypostatized conception of information. The constraint is that superluminal signalling is ruled out on pain of temporal loop paradoxes (\citet[e.g.][\S 7.ix]{rindler}). What this means is that no \textit{physical process} is permissible that would allow a signal to be sent superluminally and thus allow information to be transmitted superluminally. What are ruled out are certain types of physical processes, not, save as a metaphor, certain types of motion of information\footnote{The types of processes in question might not be identifiable without recourse to concepts of what would count as successful transmission of information, but this does not mean that one has to conceive of information as an entity or substance, just that one needs a concept of what it means to receive a signal from which one can learn something.}. 

A final objection that might be raised to support the line of thought that inclines one towards the Jozsa and Penrose conception of teleportation is just this: 
Well, don't we after all require that information be propagated in a spatio-temporally continuous way? Even if this is not to be construed as a flow of stuff, or the passage of an entity?

The response illustrates part of the value of noting the features of the term `information' that have been emphasized here. 

The genuine question we face is: what are the physical processes that may be used to transmit information? Not the (obscure) question `How does information behave?'. Once we see what the question is clearly, then the answer, surely, is to be given by our best physical theory describing the protocol in question. To be sure, many of the most familiar classical examples we are used-to use spatio-temporally continuous changes in physical properties to transmit information (a prime example might be the use of radio waves), but it is up to physical theory to tell us about the nature of the processes we are using to transmit information in any given situation. And the examples we have found in entanglement assisted communication seem precisely to be examples in which \textit{global} rather than local properties are being used to carry information; and there seems not to be a useful sense in which information is being carried in a spatio-temporally continuous way (although, see \citet{nifpaper} for further discussion of Deutsch and Hayden's opposing view).

It is not the nature of information that is at issue, but the nature of the physical objects and the physical properties we may use to transmit information.

On a final note, the deflationary approach towards teleportation that I have advocated should be compared with what may be termed the `nihilist' approach of \citet{duwell:2003}. While I am in broad sympathy with much of what Duwell has to say, we differ on some important points. Duwell also advocates the view that quantum information is not a substance, but reaches from this the strong conclusion that quantum information does not exist. From the current point of view this conclusion is unwarranted. Certainly, quantum information is not a substance or entity, but this does not mean that it doesn't exist, it is just a reflection of the fact that `information' is an abstract noun. `Beauty' for example, is an abstract noun, but no one would want to conclude that there is no beauty in the world. Moreover, Duwell's conclusion could only possibly be hyperbolical, for if classical information can be said to exist, then so too can quantum information; and contrapositively, if quantum information does not exist, then no more does classical information. The concept of classical information is given by Shannon's noiseless coding theorem, the concept of quantum information, by the quantum noiseless coding theorem. As we are by now vividly aware, these are not concepts of material quantities or things. But rejecting the concept of quantum information would be akin to cutting off one's nose to spite one's face; and is by no means necessary in order to get a proper understanding of teleportation.

Teleportation is not rendered unproblematic by trying to do without the notion of quantum information and facing the protocol equipped only with Shannon's concept, but simply by resisting the temptation to hypostatize an abstract noun; and, having  recognised the status of `information' as an abstract noun, by realising that the only genuine question one faces is the relatively straightforward one of describing the physical processes by which information is transmitted.

\section*{Acknowledgements}

Thanks are due to Jon Barrett, Harvey Brown and Peter Morgan for useful discussion, to Jane Timpson for the Figure; and to John Christie and Joesph Melia for asking some good questions.
\newpage
\appendix

\section{Elements of information theory}\label{elements}

The Shannon information measure is defined as:
\[H(X) = -\sum_{i=1}^{n} p(x_{i})\log p(x_{i}); \]
logarithms to base 2. Its primary role is in characterising the degree to which the output of a source modelled as producing letters picked from a finite alphabet $\{ x_{1},\ldots,x_{n}\}$ with probabilities $p(x_{i})$ may be compressed. \citet{shannon} showed in his noiseless coding theorem that for very long messages of length $N$ produced by such a source, it was possible to encode them onto a string of $NH(X)$ bits, thus effecting a compression from $2^{N\log n}$ (the number of possible messages of length $N$ drawn from an alphabet of $n$ letters) to $2^{NH(X)}$.

The value of $H(X)$ depends on the probability distribution $\{p(x_{i})\}$, in fact:
\[0 \leq H(X) \leq \log n,\]  
where the minimum is achieved when all but one of the $p(x_{i})$ are zero (the distribution is maximally peaked); and the maximum when the distribution is flat, $ \forall i\; p(x_{i}) = 1/n$. Thus, a source that always outputs the same letter has zero information (cf. footnote~\ref{noinfo}); and one that produces each letter with equal probability has maximum information for a given alphabet size.

$H(X)$ may also be understood as a measure of uncertainty, that is, as a measure of how concentrated the probability distribution $\{p(x_{i})\}$ is (\citet{jos}). But this r\^ole is logically independent of its r\^ole in characterising information sources (\textit{ibid.}), a fact which is rarely recognised and has accordingly given rise to considerable confusion. (\textit{Ibid.} See \citet{supposed} for discussion.) 

A \textit{communication channel} (to be precise, a \textit{discrete memoryless} channel) may be characterised in terms of conditional probabilities $p(y_{j}|x_{i})$: given that input $x_{i}$ to the channel is prepared, what is the probability of getting output $y_{j}$? If the channel is noiseless, then the mapping from input to output states will be one-to-one; in the presence of noise, a given input may give rise to a spread of outputs with various probabilities.
In his \textit{noisy coding theorem}, \citet{shannon} proved the surprising result that even in the presence of noise, it is possible to send messages over a channel with a probability of error that tends to zero as the length of message, $N$, tends to infinity. 

Given an input distribution $p(x_{i})$ and the conditional probabilities $p(y_{j}|x_{i})$ characterising the channel, define the \textit{mutual information} $H(X:Y)$:

\[H(X:Y) = H(X) - H(X|Y),\]
where the `conditional entropy' $H(X|Y) = \sum_{j}p(y_{j})\bigr(-\sum_{i} p(x_{i}|y_{j})\log p(x_{i}|y_{j})\bigr)$. It follows from the noisy coding theorem that the mutual information governs the rate at which it is possible to send information over a channel with input distribution $p(x_{i})$: It will be possible to send the output of any information source $W$ with an information $H(W)= H(X:Y)$ over the channel, with arbitrarily small error as $N$ is increased. 
The \textit{capacity} of a channel is defined as the supremum over input distributions $p(x_{i})$ of the mutual information $H(X:Y)$.

A quantum source (\citet{qcoding}) produces systems in signal states $\rho_{x_{i}}$ with probabilities $p(x_{i})$. A density operator $\rho=\sum_{i}p(x_{i})\rho_{x_{i}}$ may be associated with the output of this source. \citet{qcoding} considered quantum sources with pure signal states $\rho_{x_{i}}= \ket{x_{i}}{}\!\bra{}{x_{i}}$. The von Neumann entropy $S(\rho)$ is defined as:
\[ S(\rho) = -\mathrm{Tr} \rho\log\rho.\]
The quantum noiseless coding theorem shows that for large $N$, the output of length $N$ of a source with signal states $\rho_{x_{i}}= \ket{x_{i}}{}\!\bra{}{x_{i}}$ may be encoded onto a quantum system of $2^{NS(\rho)}$ dimensions, that is, using $NS(\rho)$ qubits.

\section{Teleportation in the absence of collapse}\label{nocollapse appendix}

The original \citet{teleportation} treatment of teleportation involved collapse following Alice's measurement in order to pick out, probabilistically, a definite state of Bob's system that depended in a fixed way on the identity of the unknown state. As we have seen, however, it is quite possible to treat the teleportation protocol in a no-collapse setting (\citet{vaidman,braunstein:irreversible,mermin:teleportation}). It might prove helpful to see how this can work.

Recall the pertinent expression of the initial state in eqn.~(\ref{rewrite2}):
\begin{multline*}
\ket{\chi}{1}\ket{\psi^{-}}{23} = \frac{1}{2}\biggl( \ket{\phi^{+}}{12} \bigl(-i\sigma_{y}^{3}\ket{\chi}{3}\bigr) + \ket{\phi^{-}}{12} \bigl(\sigma_{x}^{3}\ket{\chi}{3}\bigr) \\
 + \ket{\psi^{+}}{12} \bigl(-\sigma_{z}^{3}\ket{\chi}{3}\bigr) + \ket{\psi^{-}}{12} \bigl(-\mathbf{1}^{3}\ket{\chi}{3}\bigr)\biggr).  
\end{multline*}
Alice is going to perform her Bell-basis measurement and she will need systems to record the outcome of that measurement. As there are four possible outcomes, she may use two qubits, call them $c$ and $d$, with an indicator basis $\{\ket{0}{},\ket{1}{}\!\}$. Her measurement interaction $U_{A}$ may be chosen (in the ideal case) so that:

\begin{singlespacing}
\begin{equation}\label{measurement}
\begin{split}
\ket{\phi^{+}}{12}\ket{0}{c}\ket{0}{d} & \mapsto \ket{\phi^{+}}{12} \ket{0}{c}\ket{0}{d}, \\
\ket{\phi^{-}}{12}\ket{0}{c}\ket{0}{d} & \mapsto \ket{\phi^{-}}{12}\ket{0}{c}\ket{1}{d}, \\
\ket{\psi^{+}}{12}\ket{0}{c}\ket{0}{d} & \mapsto \ket{\psi^{+}}{12}\ket{1}{c}\ket{0}{d}, \\
\ket{\psi^{-}}{12}\ket{0}{c}\ket{0}{d} & \mapsto \ket{\psi^{-}}{12}\ket{1}{c}\ket{1}{d}.
\end{split}
\end{equation}
\end{singlespacing}

Before Alice's measurement, then, the total state is $\ket{0}{c}\ket{0}{d}\ket{\chi}{1}\ket{\psi^{-}}{23}$. Note that relative to the states of $c$ and $d$, the state of system $1$ at Alice's locale is definite, but the state of Bob's system, $3$, is not. Being maximally mixed (half of a maximally entangled pair) it has no pure state of its own.

When Alice performs her measurement on 1 and 2, the mapping (\ref{measurement}) will apply and all the systems will now become entangled. The new total state is:
  
\begin{multline}
\ket{\Psi}{\rm{tot}} = \frac{1}{2}\biggl( \ket{0}{c}\ket{0}{d}\ket{\phi^{+}}{12} \bigl(-i\sigma_{y}^{3}\ket{\chi}{3}\bigr) + \ket{0}{c}\ket{1}{d}\ket{\phi^{-}}{12} \bigl(\sigma_{x}^{3}\ket{\chi}{3}\bigr) \\
 + \ket{1}{c}\ket{0}{d}\ket{\psi^{+}}{12} \bigl(-\sigma_{z}^{3}\ket{\chi}{3}\bigr) + \ket{1}{c}\ket{1}{d}\ket{\psi^{-}}{12} \bigl(-\mathbf{1}^{3}\ket{\chi}{3}\bigr)\biggr).  
\end{multline}
We can see that system 1 is certainly no longer in the state $\ket{\chi}{}$, it is in fact now thoroughly entangled and has the reduced state $1/2\,\mathbf{1}$. But also, crucially, relative to definite measurement outcome states of systems $c$ and $d$, Bob's system now has a definite state related to the initial $\ket{\chi}{}$. As everything is entangled, though, the \textit{reduced} state of his system is still simply $1/2\,\mathbf{1}$. 

We need to disentangle system 3 from all the others; and preferably in such a way that it is left in the unknown state $\ket{\chi}{}$. To do this, Bob will need to perform a conditional unitary operation, that is, an operation on his system that depends on the states of the two qubits recording the outcome of Alice's measurement. As he can only apply operations locally (the standard assumption) systems $c$ and $d$ will need to be transported to him before the operation may be performed. 

Once $c$ and $d$ are with Bob, he may perform the following conditional unitary transformation, $U_{B}$:
\[U_{B} = P^{cd}_{\ket{00}{}}\otimes (i\sigma_{y}^{3})\, +\, P^{cd}_{\ket{01}{}}\otimes\sigma_{x}^{3}\, + \,P^{cd}_{\ket{10}{}}\otimes(-\sigma_{z}^{3})\, +\, P^{cd}_{\ket{11}{}}\otimes (-\mathbf{1}^{3}),  \]  
where $P^{cd}_{\ket{00}{}}$ is the projector onto $\ket{0}{c}\ket{0}{d}$, and so on. This will have the effect of disentangling 3 and leaving it in the unknown state $\ket{\chi}{}$, as desired (note that $U_{B}$ does not depend on the identity of $\ket{\chi}{}$):
\begin{equation}
U_{B}\ket{\Psi}{\rm{tot}} = \frac{1}{2}\biggr(\ket{\phi^{+}}{12}\ket{0}{c}\ket{0}{d} + \ket{\phi^{-}}{12}\ket{0}{c}\ket{1}{d} + \ket{\psi^{+}}{12}\ket{1}{c}\ket{0}{d} + \ket{\psi^{-}}{12}\ket{1}{c}\ket{1}{d}\biggl)\ket{\chi}{3}.
\end{equation}



\newpage \small
\begin{thebibliography}{69}
\newcommand{\enquote}[1]{`#1'}
\providecommand{\natexlab}[1]{#1}
\providecommand{\url}[1]{\texttt{#1}}
\providecommand{\urlprefix}{URL }
\expandafter\ifx\csname urlstyle\endcsname\relax
  \providecommand{\doi}[1]{doi:\discretionary{}{}{}#1}\else
  \providecommand{\doi}{doi:\discretionary{}{}{}\begingroup
  \urlstyle{rm}\Url}\fi
\providecommand{\bibinfo}[2]{#2}
\providecommand{\eprint}[2][]{\url{#2}}

\bibitem[{Alvarez and Hyman(1998)}]{hyman:alvarez}
\bibinfo{author}{Alvarez, M.} and \bibinfo{author}{Hyman, J.}
  [\bibinfo{year}{1998}]: \enquote{\bibinfo{title}{Agents and their Actions}},
  \textit{\bibinfo{journal}{Philosophy}}, \textbf{\bibinfo{volume}{73}}, pp.
  \bibinfo{pages}{219--45}.

\bibitem[{Barrett(2001)}]{jon1}
\bibinfo{author}{Barrett, J.~S.} [\bibinfo{year}{2001}]:
  \enquote{\bibinfo{title}{Implications of Teleportation for Nonlocality}},
  \textit{\bibinfo{journal}{Physical Review A}}, \textbf{\bibinfo{volume}{64}},
  p. \bibinfo{pages}{042\,305}, \bibinfo{note}{{a}r{X}iv:quant-ph/0103105}.

\bibitem[{Bell(1966)}]{bell:1966}
\bibinfo{author}{Bell, J.~S.} [\bibinfo{year}{1966}]:
  \enquote{\bibinfo{title}{On the Problem of Hidden Variables in Quantum
  Mechanics}}, \textit{\bibinfo{journal}{Reviews of Modern Physics}},
  \textbf{\bibinfo{volume}{38}}, pp. \bibinfo{pages}{447--52},
  \bibinfo{note}{repr. in \citet[Chpt.1]{bell:speakable}.}

\bibitem[{Bell(1981)}]{bell:1981}
\bibinfo{author}{Bell, J.~S.} [\bibinfo{year}{1981}]:
  \enquote{\bibinfo{title}{Quantum Mechanics for Cosmologists}}, in
  \bibinfo{editor}{C.~Isham}, \bibinfo{editor}{R.~Penrose} and
  \bibinfo{editor}{D.~Sciama} (eds.), \textit{\bibinfo{booktitle}{Quantum
  Gravity 2}}, \bibinfo{address}{Oxford}: \bibinfo{publisher}{Oxford University
  Press}, pp. \bibinfo{pages}{611--37}, \bibinfo{note}{repr. in
  \citet[Chpt.15]{bell:speakable}.}

\bibitem[{Bell(1987)}]{bell:speakable}
\bibinfo{author}{Bell, J.~S.} [\bibinfo{year}{1987}]:
  \textit{\bibinfo{title}{Speakable and Unspeakable in Quantum Mechanics}},
  \bibinfo{address}{Cambridge}: \bibinfo{publisher}{Cambridge University
  Press}.

\bibitem[{Bennett \textit{et~al.}(1993)Bennett, Brassard, Cr\'epeau, Jozsa,
  Peres and Wootters}]{teleportation}
\bibinfo{author}{Bennett, C.~H.}, \bibinfo{author}{Brassard, G.},
  \bibinfo{author}{Cr\'epeau, C.}, \bibinfo{author}{Jozsa, R.},
  \bibinfo{author}{Peres, A.} and \bibinfo{author}{Wootters, W.}
  [\bibinfo{year}{1993}]: \enquote{\bibinfo{title}{Teleporting an Unknown State
  via Dual Classical and {EPR} channels}}, \textit{\bibinfo{journal}{Physical
  Review Letters}}, \textbf{\bibinfo{volume}{70}}, pp.
  \bibinfo{pages}{1895--99}.

\bibitem[{Bohm(1951)}]{bohm}
\bibinfo{author}{Bohm, D.} [\bibinfo{year}{1951}]:
  \textit{\bibinfo{title}{Quantum Theory}}, chapter~\bibinfo{chapter}{22},
  \bibinfo{address}{Englewood Cliffs}: \bibinfo{publisher}{Prentice-Hall}, pp.
  \bibinfo{pages}{615--619}.

\bibitem[{Bohm(1952)}]{bohm:1952}
\bibinfo{author}{Bohm, D.} [\bibinfo{year}{1952}]: \enquote{\bibinfo{title}{A
  Suggested Interpretation of the Quantum Theory in Terms of Hidden Variables,
  {I} and {II}}}, \textit{\bibinfo{journal}{Physical Review}},
  \textbf{\bibinfo{volume}{85}}, pp. \bibinfo{pages}{166--79;180--93}.

\bibitem[{Bohm and Hiley(1993)}]{bohmhiley}
\bibinfo{author}{Bohm, D.} and \bibinfo{author}{Hiley, B.~J.}
  [\bibinfo{year}{1993}]: \textit{\bibinfo{title}{The Undivided Universe: An
  Ontological Interpretation of Quantum Theory}}, \bibinfo{address}{London}:
  \bibinfo{publisher}{Routledge}.

\bibitem[{Bohm \textit{et~al.}(1955)Bohm, Schiller and Tiomno}]{bst:1955}
\bibinfo{author}{Bohm, D.}, \bibinfo{author}{Schiller, R.} and
  \bibinfo{author}{Tiomno, J.} [\bibinfo{year}{1955}]:
  \enquote{\bibinfo{title}{A Classical Interpretation of the {P}auli
  Equation}}, \textit{\bibinfo{journal}{Nuovo Cimento}},
  \textbf{\bibinfo{volume}{Supp 1}}, pp. \bibinfo{pages}{48--66}.

\bibitem[{Braunstein(1996)}]{braunstein:irreversible}
\bibinfo{author}{Braunstein, S.~L.} [\bibinfo{year}{1996}]:
  \enquote{\bibinfo{title}{Quantum Teleportation without Irreversible
  Detection}}, \textit{\bibinfo{journal}{Physical Review A}},
  \textbf{\bibinfo{volume}{53}}, pp. \bibinfo{pages}{1900--2}.

\bibitem[{Brown and Wallace(2005)}]{bw:2005}
\bibinfo{author}{Brown, H.~R.} and \bibinfo{author}{Wallace, D.}
  [\bibinfo{year}{2005}]: \enquote{\bibinfo{title}{Solving the Measurement
  Problem: De {B}roglie-{B}ohm Loses Out to {E}verett}},
  \textit{\bibinfo{journal}{Foundations of Physics}},
  \textbf{\bibinfo{volume}{35}}, pp. \bibinfo{pages}{517--40},
  \bibinfo{note}{ar{X}iv:quant-ph/0403094}.

\bibitem[{Brukner and Zeilinger(2001)}]{conceptualinadequacy}
\bibinfo{author}{Brukner, C.} and \bibinfo{author}{Zeilinger, A.}
  [\bibinfo{year}{2001}]: \enquote{\bibinfo{title}{Conceptual inadequacy of the
  {S}hannon information in quantum measurements}},
  \textit{\bibinfo{journal}{Physical Review A}}, \textbf{\bibinfo{volume}{63}},
  p. \bibinfo{pages}{022\,113}.

\bibitem[{Bub(1997)}]{bub:1997}
\bibinfo{author}{Bub, J.} [\bibinfo{year}{1997}]:
  \textit{\bibinfo{title}{Interpreting the Quantum World}},
  \bibinfo{address}{Cambridge}: \bibinfo{publisher}{Cambridge University
  Press}, \bibinfo{edition}{first paperback (1999)} edition.

\bibitem[{Busch(1997)}]{busch:observable}
\bibinfo{author}{Busch, P.} [\bibinfo{year}{1997}]: \enquote{\bibinfo{title}{Is
  the quantum state (an) observable?}}, in \bibinfo{editor}{R.~S. Cohen},
  \bibinfo{editor}{M.~Horne} and \bibinfo{editor}{J.~Stachel} (eds.),
  \textit{\bibinfo{booktitle}{Potentiality, Entanglement and
  Passion-at-a-Distance}}, \bibinfo{address}{Dordrecht}:
  \bibinfo{publisher}{Kluwer Academic Pubishers}, pp. \bibinfo{pages}{61--70},
  \bibinfo{note}{ar{X}iv:quant-ph/9604014}.

\bibitem[{Clifton and Pope(2001)}]{cliftonpope}
\bibinfo{author}{Clifton, R.} and \bibinfo{author}{Pope, D.}
  [\bibinfo{year}{2001}]: \enquote{\bibinfo{title}{On the Nonlocality of the
  Quantum Channel in the Standard Teleportation Protocol}},
  \textit{\bibinfo{journal}{Physics Letters A}},
  \textbf{\bibinfo{volume}{292}}, pp. \bibinfo{pages}{1--11},
  \bibinfo{note}{ar{X}iv:quant-ph/0103075}.

\bibitem[{Davidson(1980)}]{davidson:actionsandevents}
\bibinfo{author}{Davidson, D.} [\bibinfo{year}{1980}]:
  \textit{\bibinfo{title}{Essays on Actions and Events}},
  \bibinfo{address}{Oxford}: \bibinfo{publisher}{Oxford University Press}.

\bibitem[{d'Espagnat(1976)}]{d'Espagnat}
\bibinfo{author}{d'Espagnat, B.} [\bibinfo{year}{1976}]:
  \textit{\bibinfo{title}{Conceptual Foundations of Quantum Mechanics}},
  \bibinfo{address}{Reading, Massachusetts}:
  \bibinfo{publisher}{Addison-Wesley}, \bibinfo{edition}{2nd} edition.

\bibitem[{Deutsch(1997)}]{FoR}
\bibinfo{author}{Deutsch, D.} [\bibinfo{year}{1997}]:
  \textit{\bibinfo{title}{The Fabric of Reality}}, \bibinfo{address}{London}:
  \bibinfo{publisher}{Penguin Books}.

\bibitem[{Deutsch(1999)}]{deutsch:decision}
\bibinfo{author}{Deutsch, D.} [\bibinfo{year}{1999}]:
  \enquote{\bibinfo{title}{Quantum Theory of Probability and Decisions}},
  \textit{\bibinfo{journal}{Proceedings of the Royal Society of London A}},
  \textbf{\bibinfo{volume}{455}}, pp. \bibinfo{pages}{3129--37},
  \bibinfo{note}{ar{X}iv:quant-ph/09906015}.

\bibitem[{Deutsch and Hayden(2000)}]{dh}
\bibinfo{author}{Deutsch, D.} and \bibinfo{author}{Hayden, P.}
  [\bibinfo{year}{2000}]: \enquote{\bibinfo{title}{Information Flow in
  Entangled Quantum Systems}}, \textit{\bibinfo{journal}{Proceedings of the
  Royal Society of London A}}, \textbf{\bibinfo{volume}{456}}, pp.
  \bibinfo{pages}{1759--74}, \bibinfo{note}{{a}r{X}iv:quant-ph/9906007}.

\bibitem[{Dieks(1982)}]{dieks}
\bibinfo{author}{Dieks, D.} [\bibinfo{year}{1982}]:
  \enquote{\bibinfo{title}{Communication by {EPR} Devices}},
  \textit{\bibinfo{journal}{Physics Letters A}}, \textbf{\bibinfo{volume}{92}},
  pp. \bibinfo{pages}{271--2}.

\bibitem[{Dirac(1947)}]{dirac}
\bibinfo{author}{Dirac, P. A.~M.} [\bibinfo{year}{1947}]:
  \textit{\bibinfo{title}{The Principles of Quantum Mechanics}},
  \bibinfo{address}{Oxford}: \bibinfo{publisher}{Oxford University Press},
  \bibinfo{edition}{3rd} edition.

\bibitem[{Dretske(1981)}]{dretske:1981}
\bibinfo{author}{Dretske, F.~I.} [\bibinfo{year}{1981}]:
  \textit{\bibinfo{title}{Knowledge and the Flow of Information}},
  \bibinfo{address}{Oxford}: \bibinfo{publisher}{Basil Blackwell}.

\bibitem[{Duwell(2003)}]{duwell:2003}
\bibinfo{author}{Duwell, A.} [\bibinfo{year}{2003}]:
  \enquote{\bibinfo{title}{Quantum information does not exist}},
  \textit{\bibinfo{journal}{Studies in History and Philosophy of Modern
  Physics}}, \textbf{\bibinfo{volume}{34}}, pp. \bibinfo{pages}{479--99}.

\bibitem[{Eberhard(1978)}]{eberhard}
\bibinfo{author}{Eberhard, P.~H.} [\bibinfo{year}{1978}]:
  \enquote{\bibinfo{title}{Bell's Theorem and the Different Concepts of
  Locality}}, \textit{\bibinfo{journal}{Nouvo Cimento}},
  \textbf{\bibinfo{volume}{46B}}, pp. \bibinfo{pages}{392--419}.

\bibitem[{Einstein \textit{et~al.}(1935)Einstein, Podolsky and Rosen}]{EPR}
\bibinfo{author}{Einstein, A.}, \bibinfo{author}{Podolsky, B.} and
  \bibinfo{author}{Rosen, N.} [\bibinfo{year}{1935}]:
  \enquote{\bibinfo{title}{Can Quantum Mechanical Description of Physical
  Reality be Considered Complete?}}, \textit{\bibinfo{journal}{Physical
  Review}}, \textbf{\bibinfo{volume}{47}}, pp. \bibinfo{pages}{777--80}.

\bibitem[{Everett(1957)}]{everett}
\bibinfo{author}{Everett, H., III} [\bibinfo{year}{1957}]:
  \enquote{\bibinfo{title}{``{R}elative State" Formulation of Quantum
  Mechanics}}, \textit{\bibinfo{journal}{Reviews of Modern Physics}},
  \textbf{\bibinfo{volume}{29}}, pp. \bibinfo{pages}{454--62}.

\bibitem[{Galv{\~a}o and Hardy(2003)}]{lucien:substituting}
\bibinfo{author}{Galv{\~a}o, E.~F.} and \bibinfo{author}{Hardy, L.}
  [\bibinfo{year}{2003}]: \enquote{\bibinfo{title}{Substituting a Qubit for an
  Arbitrarily Large Number of Classical Bits}},
  \textit{\bibinfo{journal}{Physical Review Letters}},
  \textbf{\bibinfo{volume}{90}}, pp. \bibinfo{pages}{087\,902--1--4}.

\bibitem[{Ghiradi \textit{et~al.}(1980)Ghiradi, Rimini and
  Weber}]{grw:no-signalling}
\bibinfo{author}{Ghiradi, G.~C.}, \bibinfo{author}{Rimini, A.} and
  \bibinfo{author}{Weber, T.} [\bibinfo{year}{1980}]:
  \enquote{\bibinfo{title}{A General Argument against Superluminal Transmission
  through the Quantum Mechanical Measurement Process}},
  \textit{\bibinfo{journal}{Lettere Nuovo Cimento}},
  \textbf{\bibinfo{volume}{24}}, pp. \bibinfo{pages}{293--8}.

\bibitem[{Ghirardi \textit{et~al.}(1986)Ghirardi, Rimini and Weber}]{grw}
\bibinfo{author}{Ghirardi, G.~C.}, \bibinfo{author}{Rimini, A.} and
  \bibinfo{author}{Weber, T.} [\bibinfo{year}{1986}]:
  \enquote{\bibinfo{title}{Unified Dynamics for Microscopic and Macroscopic
  Systems}}, \textit{\bibinfo{journal}{Physical Review D}},
  \textbf{\bibinfo{volume}{34}}, pp. \bibinfo{pages}{470--91}.

\bibitem[{Hardy(1999)}]{hardy:disentangling}
\bibinfo{author}{Hardy, L.} [\bibinfo{year}{1999}]:
  \enquote{\bibinfo{title}{Disentangling Nonlocality and Teleportation}},
  \bibinfo{note}{ar{X}iv:quant-ph/9906123}.

\bibitem[{Hewitt-Horsman(2002)}]{horsman}
\bibinfo{author}{Hewitt-Horsman, C.} [\bibinfo{year}{2002}]:
  \enquote{\bibinfo{title}{Quantum Computation and Many Worlds}},
  \bibinfo{note}{ar{X}iv:quant-ph/0210204}.

\bibitem[{Hiley(1999)}]{hiley:1999}
\bibinfo{author}{Hiley, B.~J.} [\bibinfo{year}{1999}]:
  \enquote{\bibinfo{title}{Active Information and Teleportation}}, in
  \bibinfo{editor}{D.~Greenberger}, \bibinfo{editor}{W.~L. Reiter} and
  \bibinfo{editor}{A.~Zeilinger} (eds.),
  \textit{\bibinfo{booktitle}{Epistemological and Experimental Perspectives on
  Quantum Physics}}, Vienna Circle Institute Yearbook,
  \bibinfo{address}{Dordrecht}: \bibinfo{publisher}{Kluwer}, pp.
  \bibinfo{pages}{113--25}.

\bibitem[{Holevo(1973)}]{holevo}
\bibinfo{author}{Holevo, A.~S.} [\bibinfo{year}{1973}]:
  \enquote{\bibinfo{title}{Information theoretical aspects of quantum
  measurement}}, \textit{\bibinfo{journal}{Problems of Information Transmission
  (USSR)}}, \textbf{\bibinfo{volume}{9}}, pp. \bibinfo{pages}{177--83}.

\bibitem[{Holland(1995)}]{holland:1995}
\bibinfo{author}{Holland, P.~R.} [\bibinfo{year}{1995}]:
  \textit{\bibinfo{title}{The Quantum Theory of Motion: An Account of the de
  {B}roglie-{B}ohm Causal Interpretation of Quantum Mechanics}},
  \bibinfo{address}{Cambridge}: \bibinfo{publisher}{Cambridge University
  Press}, \bibinfo{edition}{first paperback} edition.

\bibitem[{Jozsa(1998)}]{jozsa:1998}
\bibinfo{author}{Jozsa, R.} [\bibinfo{year}{1998}]:
  \enquote{\bibinfo{title}{Entanglement and Quantum Computation}}, in
  \bibinfo{editor}{S.~Huggett}, \bibinfo{editor}{L.~Mason},
  \bibinfo{editor}{K.~P. Tod}, \bibinfo{editor}{S.~T. Tsou} and
  \bibinfo{editor}{N.~M.~J. Woodhouse} (eds.), \textit{\bibinfo{booktitle}{The
  Geometric Universe}}, \bibinfo{address}{Oxford}: \bibinfo{publisher}{Oxford
  University Press}, pp. \bibinfo{pages}{369--379},
  \bibinfo{note}{ar{X}iv:quant-ph/9707034}.

\bibitem[{Jozsa(2004)}]{jozsa:2003}
\bibinfo{author}{Jozsa, R.} [\bibinfo{year}{2004}]:
  \enquote{\bibinfo{title}{Illustrating the concept of quantum information}},
  \textit{\bibinfo{journal}{IBM Journal of Research and Development}},
  \textbf{\bibinfo{volume}{4}}, pp. \bibinfo{pages}{79--85},
  \bibinfo{note}{ar{X}iv:quant-ph/0305114}.

\bibitem[{Jozsa and Schumacher(1994)}]{schumacher:jozsa}
\bibinfo{author}{Jozsa, R.} and \bibinfo{author}{Schumacher, B.}
  [\bibinfo{year}{1994}]: \enquote{\bibinfo{title}{A New Proof of the Quantum
  Noiseless Coding Theorem}}, \textit{\bibinfo{journal}{Journal of Modern
  Optics}}, \textbf{\bibinfo{volume}{41}}, pp. \bibinfo{pages}{2343--49}.

\bibitem[{Maroney and Hiley(1999)}]{maroney:hiley}
\bibinfo{author}{Maroney, O.} and \bibinfo{author}{Hiley, B.}
  [\bibinfo{year}{1999}]: \enquote{\bibinfo{title}{Quantum State Teleportation
  Understood Through the {B}ohm Interpretation}},
  \textit{\bibinfo{journal}{Foundations of Physics}},
  \textbf{\bibinfo{volume}{29}}, pp. \bibinfo{pages}{1403--15}.

\bibitem[{Mermin(2001)}]{mermin:teleportation}
\bibinfo{author}{Mermin, N.~D.} [\bibinfo{year}{2001}]:
  \enquote{\bibinfo{title}{From Classical State-Swapping to Teleportation}},
  \textit{\bibinfo{journal}{Physical Review A}}, \textbf{\bibinfo{volume}{65}},
  p. \bibinfo{pages}{012\,320}, \bibinfo{note}{ar{X}iv:quant-ph/0105117}.

\bibitem[{Morgan(2001)}]{petermorgan}
\bibinfo{author}{Morgan, P.} [\bibinfo{year}{2001}]: \bibinfo{note}{Personal
  communication}.

\bibitem[{Park(1970)}]{park:1970}
\bibinfo{author}{Park, J.~L.} [\bibinfo{year}{1970}]:
  \enquote{\bibinfo{title}{The Concept of Transition in Quantum Mechanics}},
  \textit{\bibinfo{journal}{Foundations of Physics}},
  \textbf{\bibinfo{volume}{1}}, p.~\bibinfo{pages}{23}.

\bibitem[{Penrose(1998)}]{penrose:1998}
\bibinfo{author}{Penrose, R.} [\bibinfo{year}{1998}]:
  \enquote{\bibinfo{title}{Quantum Computation, Entanglement and State
  Reduction}}, \textit{\bibinfo{journal}{Philosophical Transactions of the
  Royal Society of London A}}, \textbf{\bibinfo{volume}{356}}, pp.
  \bibinfo{pages}{1927--39}.

\bibitem[{Redhead(1987)}]{redhead:no-signalling}
\bibinfo{author}{Redhead, M. L.~G.} [\bibinfo{year}{1987}]:
  \textit{\bibinfo{title}{Incompleteness, Non-Locality and Realism}}, chapter
  \bibinfo{chapter}{4.6}, \bibinfo{address}{Oxford}: \bibinfo{publisher}{Oxford
  University Press}.

\bibitem[{Rindler(1991)}]{rindler}
\bibinfo{author}{Rindler, W.} [\bibinfo{year}{1991}]:
  \textit{\bibinfo{title}{Introduction to Special Relativity}},
  \bibinfo{address}{Oxford}: \bibinfo{publisher}{Oxford University Press},
  \bibinfo{edition}{2nd} edition.

\bibitem[{Saunders(1995)}]{simon1}
\bibinfo{author}{Saunders, S.} [\bibinfo{year}{1995}]:
  \enquote{\bibinfo{title}{Time, Quantum Mechanics, and Decoherence}},
  \textit{\bibinfo{journal}{Synthese}}, \textbf{\bibinfo{volume}{102}}, pp.
  \bibinfo{pages}{235--66}.

\bibitem[{Saunders(1996{\natexlab{a}})}]{simonrelativism}
\bibinfo{author}{Saunders, S.} [\bibinfo{year}{1996}{\natexlab{a}}]:
  \enquote{\bibinfo{title}{Relativism}}, in \bibinfo{editor}{R.~Clifton} (ed.),
  \textit{\bibinfo{booktitle}{Perspectives on Quantum Reality}},
  \bibinfo{address}{Dordrecht}: \bibinfo{publisher}{Kluwer Academic
  Publishers}, pp. \bibinfo{pages}{125--42}.

\bibitem[{Saunders(1996{\natexlab{b}})}]{simon2}
\bibinfo{author}{Saunders, S.} [\bibinfo{year}{1996}{\natexlab{b}}]:
  \enquote{\bibinfo{title}{Time, Quantum Mechanics, and Tense}},
  \textit{\bibinfo{journal}{Synthese}}, \textbf{\bibinfo{volume}{107}}, pp.
  \bibinfo{pages}{19--53}.

\bibitem[{Saunders(1998)}]{simon3}
\bibinfo{author}{Saunders, S.} [\bibinfo{year}{1998}]:
  \enquote{\bibinfo{title}{Time, Quantum Mechanics, and Probability}},
  \textit{\bibinfo{journal}{Synthese}}, \textbf{\bibinfo{volume}{114}}, pp.
  \bibinfo{pages}{373--404}.

\bibitem[{Schumacher(1995)}]{qcoding}
\bibinfo{author}{Schumacher, B.} [\bibinfo{year}{1995}]:
  \enquote{\bibinfo{title}{Quantum Coding}}, \textit{\bibinfo{journal}{Physical
  Review A}}, \textbf{\bibinfo{volume}{51}}, pp. \bibinfo{pages}{2738--47}.

\bibitem[{Shannon(1948)}]{shannon}
\bibinfo{author}{Shannon, C.~E.} [\bibinfo{year}{1948}]:
  \enquote{\bibinfo{title}{The mathematical theory of communication}},
  \textit{\bibinfo{journal}{Bell Systems Technical Journal}},
  \textbf{\bibinfo{volume}{27}}, pp. \bibinfo{pages}{379--423, 623--56},
  \bibinfo{note}{repr. in \cite{shannon:weaver} pp.30-125}.

\bibitem[{Shannon and Weaver(1963)}]{shannon:weaver}
\bibinfo{author}{Shannon, C.~E.} and \bibinfo{author}{Weaver, W.}
  [\bibinfo{year}{1963}]: \textit{\bibinfo{title}{The Mathematical Theory of
  Communication}}, \bibinfo{address}{Urbana and Chicago}:
  \bibinfo{publisher}{University of Illinois Press}, \bibinfo{edition}{{I}llini
  {P}ress} edition.

\bibitem[{Shimony(1984)}]{shimony}
\bibinfo{author}{Shimony, A.} [\bibinfo{year}{1984}]:
  \enquote{\bibinfo{title}{Controllable and Uncontrollable Non-Locality}}, in
  \bibinfo{editor}{S.~Kamefuchi} (ed.), \textit{\bibinfo{booktitle}{Foundations
  of Quantum Mechanics in the Light of New Technology}},
  \bibinfo{address}{Tokyo}: \bibinfo{publisher}{The Physical Society of Japan},
  \bibinfo{note}{repr. in A. Shimony, 1993 \textit{Search for a Naturalistic
  World View}, Vol. 2, Cambridge: Cambridge University Press, pp.130-139.}

\bibitem[{Steane(2003)}]{steane}
\bibinfo{author}{Steane, A.~M.} [\bibinfo{year}{2003}]:
  \enquote{\bibinfo{title}{A quantum computer only needs one universe}},
  \textit{\bibinfo{journal}{Studies in History and Philosophy of Modern
  Physics}}, \textbf{\bibinfo{volume}{34}}, pp. \bibinfo{pages}{469--78}.

\bibitem[{Tausk(1967)}]{tausk}
\bibinfo{author}{Tausk, K.} [\bibinfo{year}{1967}]:
  \textit{\bibinfo{title}{Measurement in Quantum Mechanics}}, Ph.D. thesis,
  \bibinfo{school}{University of S\~ao Paulo}, \bibinfo{note}{pp.29-31}.

\bibitem[{Timpson(2003)}]{supposed}
\bibinfo{author}{Timpson, C.~G.} [\bibinfo{year}{2003}]:
  \enquote{\bibinfo{title}{On a Supposed Conceptual Inadequacy of the {S}hannon
  Information in Quantum Mechanics}}, \textit{\bibinfo{journal}{Studies in
  History and Philosophy of Modern Physics}}, \textbf{\bibinfo{volume}{34}},
  pp. \bibinfo{pages}{441--68}.

\bibitem[{Timpson(2004)}]{thesis}
\bibinfo{author}{Timpson, C.~G.} [\bibinfo{year}{2004}]:
  \textit{\bibinfo{title}{Quantum Information Theory and the Foundations of
  Quantum Mechanics}}, Ph.D. thesis, \bibinfo{school}{University of Oxford},
  \bibinfo{note}{ar{X}iv:quant-ph/0412063}.

\bibitem[{Timpson(2005)}]{nifpaper}
\bibinfo{author}{Timpson, C.~G.} [\bibinfo{year}{2005}]:
  \enquote{\bibinfo{title}{Nonlocality and Information Flow: The Approach of
  {D}eutsch and {H}ayden}}, \textit{\bibinfo{journal}{Foundations of Physics}},
  \textbf{\bibinfo{volume}{35}}, pp. \bibinfo{pages}{313--43},
  \bibinfo{note}{ar{X}iv:quant-ph/0312155}.

\bibitem[{Timpson and Brown(2002)}]{erpart1}
\bibinfo{author}{Timpson, C.~G.} and \bibinfo{author}{Brown, H.~R.}
  [\bibinfo{year}{2002}]: \enquote{\bibinfo{title}{Entanglement and
  Relativity}}, in \bibinfo{editor}{R.~Lupacchini} and
  \bibinfo{editor}{V.~Fano} (eds.), \textit{\bibinfo{booktitle}{Understanding
  Physical Knowledge}}, \bibinfo{address}{Bologna}:
  \bibinfo{publisher}{University of Bologna, CLUEB}, pp.
  \bibinfo{pages}{147--66}, \bibinfo{note}{ar{X}iv:quant-ph/0212140}.

\bibitem[{Timpson and Brown(2004)}]{impsep}
\bibinfo{author}{Timpson, C.~G.} and \bibinfo{author}{Brown, H.~R.}
  [\bibinfo{year}{2004}]: \enquote{\bibinfo{title}{Proper and Improper
  Separability}}, \bibinfo{note}{ar{X}iv:quant-ph/0402094, forthcoming in
  \textit{International Journal of Quantum Information}}.

\bibitem[{Uffink(1990)}]{jos}
\bibinfo{author}{Uffink, J.} [\bibinfo{year}{1990}]:
  \textit{\bibinfo{title}{Measures of Uncertainty and the Uncertainty
  Principle}}, Ph.D. thesis, \bibinfo{school}{University of Amsterdam}.

\bibitem[{Vaidman(1994)}]{vaidman}
\bibinfo{author}{Vaidman, L.} [\bibinfo{year}{1994}]:
  \enquote{\bibinfo{title}{On the Paradoxical Aspects of New Quantum
  Experiments}}, in \bibinfo{editor}{D.~Hull}, \bibinfo{editor}{M.~Forbes} and
  \bibinfo{editor}{R.~Burian} (eds.), \textit{\bibinfo{booktitle}{PSA 1994}},
  volume~\bibinfo{volume}{1}, \bibinfo{publisher}{Philosophy of Science
  Association}, pp. \bibinfo{pages}{211--7}.

\bibitem[{von Neumann(1955)}]{vN}
\bibinfo{author}{von Neumann, J.} [\bibinfo{year}{1955}]:
  \textit{\bibinfo{title}{The Mathematical Foundations of Quantum Mechanics}},
  \bibinfo{address}{Princeton}: \bibinfo{publisher}{Princeton University
  Press}, \bibinfo{note}{English translation}.

\bibitem[{Wallace(2002)}]{wallace:worlds}
\bibinfo{author}{Wallace, D.} [\bibinfo{year}{2002}]:
  \enquote{\bibinfo{title}{Worlds in the {E}verett Interpretation}},
  \textit{\bibinfo{journal}{Studies in History and Philosophy of Modern
  Physics}}, \textbf{\bibinfo{volume}{33}}, pp. \bibinfo{pages}{637--61},
  \bibinfo{note}{{a}r{X}iv:quant-ph/0103092}.

\bibitem[{Wallace(2003{\natexlab{a}})}]{wallace:structure}
\bibinfo{author}{Wallace, D.} [\bibinfo{year}{2003}{\natexlab{a}}]:
  \enquote{\bibinfo{title}{Everett and Structure}},
  \textit{\bibinfo{journal}{Studies in History and Philosophy of Modern
  Physics}}, \textbf{\bibinfo{volume}{34}}, pp. \bibinfo{pages}{87--105}.

\bibitem[{Wallace(2003{\natexlab{b}})}]{wallace:rationality}
\bibinfo{author}{Wallace, D.} [\bibinfo{year}{2003}{\natexlab{b}}]:
  \enquote{\bibinfo{title}{Everettian rationality: Defending {D}eutsch's
  approach to probability in the {E}verett interpretation}},
  \textit{\bibinfo{journal}{Studies in History and Philosophy of Modern
  Physics}}, \textbf{\bibinfo{volume}{34}}, pp. \bibinfo{pages}{415--40}.

\bibitem[{Wittgenstein(1958)}]{wittgenstein:blue}
\bibinfo{author}{Wittgenstein, L.} [\bibinfo{year}{1958}]:
  \textit{\bibinfo{title}{The Blue and Brown Books}},
  \bibinfo{address}{Oxford}: \bibinfo{publisher}{Blackwell}.

\bibitem[{Wootters and Zurek(1982)}]{wootters:zurek}
\bibinfo{author}{Wootters, W.~K.} and \bibinfo{author}{Zurek, W.~H.}
  [\bibinfo{year}{1982}]: \enquote{\bibinfo{title}{A Single Quantum Cannot be
  Cloned}}, \textit{\bibinfo{journal}{Nature}}, \textbf{\bibinfo{volume}{299}},
  pp. \bibinfo{pages}{802--3}.

\end{thebibliography}
\end{document}